\documentclass[12pt]{article}
\pdfoutput=1

\newlength{\abstractwidth}
\usepackage{amsmath}
\usepackage{amsfonts}
\usepackage{amssymb}

\usepackage{epsf}
\usepackage{color}
\usepackage{graphicx}
\usepackage{tikz}
\usepackage{dsfont}

\usepackage{hyperref}
\definecolor{darkred}{rgb}{0.8,0.1,0.1}
\hypersetup{colorlinks=true, linkcolor=darkred, citecolor=blue, linktoc=page}

\makeatletter\def\l@subsubsection#1#2{}%
\makeatother

\usepackage{latexsym}

\renewcommand{\thefootnote}{\fnsymbol{footnote}}
\renewcommand{\thanks}[1]{\footnote{#1}}
\newcommand{\starttext}{
\setcounter{footnote}{0}
\renewcommand{\thefootnote}{\arabic{footnote}}}

\newcommand{\bea}{\begin{eqnarray}}
\newcommand{\eea}{\end{eqnarray}}
\newcommand{\be}{\begin{eqnarray}}
\newcommand{\ee}{\end{eqnarray}}

\def\cA{{\cal A}}
\def\cB{{\cal B}}
\def\cC{{\cal C}}

\def\cG{{\cal G}}

\def\cI{{\cal I}}
\def\cJ{{\cal J}}
\def\cK{{\cal K}}
\def\cL{{\cal L}}

\def\cN{{\cal N}}
\def\cO{{\cal O}}
\def\cP{{\cal P}}

\def\cT{{\cal T}}

\def\mg{\mathfrak{g}}

\def\ZZ{{\mathbb Z}}
\def\RR{{\mathbb R}}

\def\CC{{\mathbb C}}

\def\Im{{\rm Im \,}}

\def\half{{1\over 2}}
\def\p{\partial}

\def\CB{C_{(2)}}

\def\a{\alpha}

\def\g{\gamma}
\def\G{\Gamma}

\def\f{\varphi}
\def\tet{\vartheta}
\def\ep{\varepsilon}

\def\pbw{\p_{\bar w}}

\def\no{\nonumber}
\def\sm{\smallskip}

\begin{document}
\starttext
\setcounter{footnote}{0}

\begin{flushright}
2016 June 1
\end{flushright}

\vskip 0.3in

\begin{center}

{\Large \bf Warped $AdS_6\times S^2$ in Type  IIB supergravity  I }

\vskip 0.1in 

{\bf \large  Local solutions}

\vskip 0.3in

{\large  Eric D'Hoker$^a$, Michael Gutperle$^a$, Andreas Karch$^b$, Christoph\,F.\,Uhlemann$^b$ }

\vskip 0.2in

{ \sl $^a$ Department of Physics and Astronomy }\\
{\sl University of California, Los Angeles, CA 90095-1547, USA}

\vskip 0.1in

{ \sl $^b$ Department of Physics} \\
{\sl  University of Washington, Seattle, WA 98195-1560, USA}\\

\vskip 0.1in

{\tt \small dhoker@physics.ucla.edu; gutperle@physics.ucla.edu;}

{\tt \small akarch@uw.edu;   uhlemann@uw.edu}

\vskip 0.3in

\begin{abstract}

\setlength{\baselineskip}{16pt}

We investigate the existence of solutions with 16 residual supersymmetries to Type IIB supergravity on a space-time of the form $AdS_6 \times S^2$ warped over a two-dimensional Riemann surface $\Sigma$. The $SO(2,5) \times SO(3)$ isometry extends to invariance under the exceptional Lie superalgebra $F(4)$. In the present paper, we construct the general Ansatz compatible with these symmetries, derive the corresponding reduced BPS equations, and obtain their complete local solution in terms of two locally holomorphic functions $\cA_\pm$ on $\Sigma$, subject to certain positivity and regularity conditions. Globally, $(\cA_+, \cA_-)$ are allowed to be multiple-valued on $\Sigma$ and be holomorphic sections of a holomorphic  bundle over $\Sigma$ with structure group contained in $SU(1,1) \times \CC$.  Globally regular solutions are expected to provide the near-horizon geometry of $(p,q)$ 5-brane and 7-brane webs which are  holographic duals to five-dimensional conformal field theories. A preliminary analysis of the positivity and regularity 
conditions will be presented here, leaving the construction of globally regular solutions to a subsequent paper.

\end{abstract}
\end{center}

\newpage

\tableofcontents

\newpage

\baselineskip=15pt
\setcounter{equation}{0}
\setcounter{footnote}{0}

\newpage

\section{Introduction}

Gauge-gravity duality, namely the equivalence between a quantum field theory in $d$ dimensions and a gravitational theory in $d+1$ dimensions via holography, has become one of the cornerstones of modern theoretical physics. The best understood examples are provided by the AdS/CFT correspondence and  involve conformal field theories (CFTs) with a large number of fields and their anti-de Sitter (AdS) dual space-times with large degrees of supersymmetry. Maldacena's original paper \cite{Maldacena:1997re} contained examples of such equivalences in $d=2,3,4$ and $6$ and many more dual pairs have been uncovered in these dimensions since then. However, what is almost entirely missing to date is any well understood example of the AdS/CFT correspondence in dimension $d=5$, namely between a 4+1 dimensional CFT and its 5+1 dimensional dual AdS space-time.

\sm

One reason that $d=5$ is special is the absence of maximally supersymmetric theories, in which the 16 Poincar\'e supersymmetries of a  4+1 dimensional CFT would be enhanced by 16 conformal supersymmetries to the maximum allowed number of 32 supersymmetries. The complete classification of superconformal algebras  \cite{Nahm:1977tg} indeed shows that $d=5$ is singled out and does not support a maximally supersymmetric CFT. Unlike its lower dimensional cousins, supersymmetric Yang-Mills theory with maximal Poincar\'e supersymmetry in 4+1 dimensions is not a conformal field theory at the origin of its moduli space. Instead, the theory is believed  to flow in the ultraviolet (UV) to the 5+1 dimensional  $(2,0)$ theory, a theory which itself features prominently in AdS/CFT (see for example \cite{Seiberg:1997ax}).

\sm

What does exist, however, are supersymmetric CFTs (SCFTs) in 4+1 dimensions with 8 Poincar\'e supersymmetries which are enhanced by 8 conformal supersymmetries to a total of 16 supersymmetries. The superconformal algebra in this case is based on the exceptional Lie superalgebra $F(4)$. This Lie superalgebra has a real form whose maximal bosonic subalgebra is $SO(2,5) \oplus SO(3)$ \cite{Minwalla:1997ka}, which may be viewed as the direct sum of the conformal algebra in 4+1 dimensions \cite{Minwalla:1997ka} and an R-symmetry. Explicit field theory examples realizing this algebra have been uncovered in \cite{Seiberg:1996bd,Intriligator:1997pq}. They are based on supersymmetric gauge theories which flow to a strongly coupled CFT  in the UV, as long as the number of matter fields coupled to the gauge field remains sufficiently small. The upper bound on the number of matter fields depends on the gauge group and on the representations of the matter content and can generically be determined from the 1-loop running of the gauge coupling \cite{Intriligator:1997pq}. Since these supersymmetric theories allow for a large $N$ limit, one 
may expect that they will possess a holographic $AdS_6$ dual.

\sm

String theory realizations of these 4+1 dimensional CFTs via brane embeddings are known and can serve as a natural starting point to construct their $AdS_6$ duals. For example, 4+1 dimensional SCFTs based on the gauge group $Sp(N)$, with hypermultiplets in the fundamental and anti-symmetric tensor representations of $Sp(N)$, were realized  in \cite{Seiberg:1996bd} via a Type IIA string theory construction involving a stack of $N$ D4 branes near a collection of $O8$ planes and $D8$ branes. A Type IIA supergravity solution based on this construction was given in \cite{Brandhuber:1999np} and generalized to quiver gauge groups in \cite{Bergman:2012kr}. While some interesting questions can be addressed in this geometry, its construction leads to singularities, because the presence of the O8 planes forces  the dilaton to blow.

\sm

A much more general brane realization of these 4+1 dimensional CFTs can be given via $(p,q)$ brane webs \cite{Aharony:1997ju,Aharony:1997bh} in Type IIB string theory. In these brane webs, D5 branes are suspended between NS5 branes, giving a 4+1 dimensional version of the construction pioneered  in \cite{Hanany:1996ie} for 2+1 dimensional gauge theories.
The brane webs can realize the gauge theory at any point on its moduli space, as well as  in the presence of relevant deformations such as mass terms or a finite bare coupling. In the limit when all the branes essentially lie on top of one another the webs do realize large classes of the 4+1 dimensional SCFTs of \cite{Intriligator:1997pq}, most naturally those based on $SU(N)$ gauge theories, possibly with couplings to matter in the fundamental representation.\footnote{Alternatively these SCFTs can be realized in terms of M-theory compactifications on singular Calabi-Yau (CY) 3-folds \cite{Intriligator:1997pq}. For the special case of toric CYs, this construction is T-dual to the $(p,q)$ web \cite{Leung:1997tw}. In any case, the realization in terms of branes serves as a more natural starting point for holographic considerations, where field theory and AdS gravity are seen as two equivalent descriptions of a given brane setup.}
D7-branes can be added to the $(p,q)$ webs \cite{DeWolfe:1999hj}, thereby slightly expanding the set of SCFTs that can be realized. As we will see, the most general supergravity Ansatz respecting the symmetries of the $(p,q)$ web does allow for a non-trivial axion winding which is required to accommodate D7-branes. When no D7-branes are present, the $(p,q)$ webs form a special sub-class for which the axion winding vanishes.

\sm

It is our goal to construct Type IIB supergravity solutions which are holographically dual to the 4+1-dimensional CFTs  realized via $(p,q)$ webs in the large $N$ limit. This will require solving the difficult problem of obtaining fully localized solutions for the corresponding intersecting branes. Examples where such solutions can give rise to warped AdS spaces were given in \cite{Youm:1999zs,Cvetic:2000cj}. Our approach will be to use the general Ansatz for the Type IIB fields consistent with the symmetries, reduce the BPS equations to this Ansatz,  and then solve the BPS equations explicitly. 

\sm

Earlier attempts at using the BPS equations appeared in \cite{Apruzzi:2014qva,Kim:2015hya,Kim:2016rhs}, where reduced BPS equations were obtained, but not generally solved. A Type IIB T-dual configuration of the D4/D8 solution in Type IIA  \cite{Brandhuber:1999np} was used to test these equations. This T-dual  solution is even more troublesome than the original Type IIA solution. In addition to the singularity caused by the presence of the O8 plane on the IIA side, one now has an additional singularity due to T-dualizing a $U(1)$ subgroup of the $SO(4)$ isometry associated with an internal sphere. The circle that is being T-dualized shrinks to zero size at the poles of the sphere, giving rise to a new singularity in the Type  IIB solution. Nevertheless, this T-dual to the D4/D8 system provides a useful check on the BPS equations obtained in \cite{Apruzzi:2014qva,Kim:2015hya,Kim:2016rhs} and we will use it for a similar purpose.

\sm

In the present paper, we will construct the general Ansatz  in Type IIB consistent with the $SO(2,5) \times SO(3)$ symmetries, derive the reduced BPS equations, and construct their general {\sl local solutions} in terms of two locally holomorphic\footnote{Throughout, we will use the terminology {\sl locally holomorphic} to include the possibility that $\cA_\pm$ may be multiple-valued on $\Sigma$, and/or have poles or other singularities on the boundary of $\Sigma$.} functions $\cA_\pm$ on a Riemann surface $\Sigma$.  To connect these local supergravity solutions to the CFTs originating from the $(p,q)$ brane system requires that we impose the necessary physical regularity conditions on the supergravity fields of the solutions. For this purpose, $\Sigma$ must be compact, with or without boundary, and geodesicly complete. The conditions on $\cA_\pm$ needed to guarantee the proper Minkowskian signature of the metric are local on $\Sigma$ and given by the following inequalities,
\bea
\label{conds}
0 & < & - |\p_w \cA_+|^2 + |\p_w \cA_-|^2
\no \\
0 & < & |\cA_+|^2 - |\cA_-|^2 + \cB + \bar \cB
\eea
where $\p_w $ is the derivative on $\Sigma$ with respect to a local holomorphic coordinate $w$, and $\cB$ is defined, up to an additive constant, by the relation $\p_w \cB = \cA_+ \p_w \cA_- - \cA_- \p_w \cA_+$. As a result, $\cB$ is also locally holomorphic. If $\Sigma$ has a boundary then the two inequalities of (\ref{conds}) must hold strictly in the interior of $\Sigma$, and become equalities on the boundary of $\Sigma$.

\sm

The group $SU(1,1)$ acting linearly on the doublet $(\cA_+, \cA_-)$ leaves the conditions (\ref{conds}) invariant and induces the standard $SU(1,1)=SL(2,\RR)$ duality transformations on the supergravity fields. 
Additionally, the supergravity solutions are invariant under constant shifts $\cA_\pm \to \cA_\pm + a_\pm$ for constants satisfying $a_-= \bar a_+$, which form the additive group isomorphic to $\CC$. The $SL(2,\ZZ)$ duality symmetry of Type IIB string theory allows us to consider supergravity solutions with identifications under $SL(2,\ZZ)$, namely with non-trivial axion winding number. Mathematically, the problem then becomes to obtain holomorphic sections $(\cA_+, \cA_-)$ of a holomorphic bundle over $\Sigma$ with structure group contained in $SL(2,\ZZ) \times \CC$, subject to the positivity conditions of (\ref{conds}). 

\sm

In the simplest case where the Riemann surface $\Sigma$ is compact and has no boundary, the second inequality in (\ref{conds}) becomes trivial since the arbitrary constant in $\cB$ can always be chosen to satisfy the inequality. The associated mathematical problem then also simplifies, and may be formulated directly in terms of a  holomorphic bundle of one-forms $\p_w \cA_\pm$ satisfying the first relation in (\ref{conds}). 
In this paper, we will provide a preliminary analysis into the existence of such global solutions but leave a detailed investigation for future work.

\sm

We close the introduction with some remarks on the relation of this work to other investigations into half-BPS solutions to Type IIB,  M-theory, and six-dimensional supergravities on space-times built as products of an $AdS$ space and one or several spheres warped over a Riemann surface~$\Sigma$. In each case, the isometries of the $AdS$ space and the sphere factors are used to reduce the BPS equations to a complicated set of non-linear  partial differential equations on~$\Sigma$,  which can be solved exactly in terms of harmonic or holomorphic data on~$\Sigma$.
This strategy was employed successfully to the construction of a large variety of novel supergravity solutions in different contexts. Type IIB supergravity duals to four-dimensional $\cN=4$ super-Yang Mills theory were found in the presence of a planar interface \cite{D'Hoker:2007xy} giving supersymmetric Janus field theories \cite{Clark:2004sb,D'Hoker:2006uv} and in the presence of Wilson loops \cite{D'Hoker:2007fq}. M-theory duals to field theories in three and six dimensions were found in the presence of various defect operators in \cite{D'Hoker:2008wc,Estes:2012vm,Bachas:2013vza}. Finally, six-dimensional supergravity duals were found to  two-dimensional conformal field theories with string junctions in \cite{Chiodaroli:2011nr}.\footnote{See also \cite{Yamaguchi:2006te,Gomis:2006sb,Gomis:2006cu,Lunin:2006xr,Lunin:2007ab} for related work on these systems using a variety of different strategies.} 

\sm

The unifying principle of this strategy was explained in \cite{D'Hoker:2014kfa} as follows. The integrability conditions on the BPS equations  produce Bianchi and field equations. With enough supersymmetry, all the Bianchi and field equations may be obtained as integrability conditions on the system of BPS equations, which therefore play a role somewhat analogous to that of a Lax pair for integrable systems. Upon reduction to $AdS_6 \times S^2$ warped over $\Sigma$, the BPS equations reduce to equations for functions on $\Sigma$ and genuinely become a set of Lax equations for a system that must therefore be integrable in the classic sense. The conformal invariance of these systems  lies at the origin of their solvability in terms of harmonic and holomorphic data on a Riemann surface $\Sigma$. 

\sm

We note that close cousins to the half-BPS solutions to Type IIB supergravity obtained here are the half-BPS solutions on a space-time of the form $AdS_2 \times S^6$ warped over a Riemann surface $\Sigma$.  The superconformal algebra is now a different real form of the exceptional Lie superalgebra $F(4)$, this time with maximal bosonic subalgebra $SO(2,1) \times SO(7)$, and 16 supersymmetries as well. The two problems are related just as duals to Wilson loops in \cite{D'Hoker:2007fq} are related to the  planar interface solutions  in  \cite{D'Hoker:2007xy}.

\subsection{Organization}

The outline of this paper is as follows. In section 2 we will review the basics of Type IIB supergravity and introduce a suitable Ansatz. In section 3 we use the decomposition of the supersymmetry generators onto Killing spinors of $AdS_6$ and $S^2$ to reduce the BPS equations for the reduced supergravity fields to a system of algebraic and partial differential equations on the surface $\Sigma$, which we partially solve.  In section 4 we solve the reduced BPS equations completely and obtain the most general local solutions to the BPS equation in terms of holomorphic data. In section 5 we summarize the expressions for all the supergravity fields in terms of the holomorphic data, analyze their behavior under $SU(1,1)$ symmetry of Type IIB supergravity, obtain the regularity conditions, recover the singular T-dual of the D4/D8 system, and conclude the section with arguments in favor of monodromy of the holomorphic data. We conclude with a discussion  in section 6.  In appendix A, a basis for the Dirac-Clifford 
algebra adapted to our Ansatz is presented,  while the geometry of Killing spinors is reviewed in appendix B. Details of the derivation of the BPS equations is in appendix C, of the Bianchi identities in appendix D, and of the expressions for the supergravity fields in terms of holomorphic data is in appendix E.

%%%%%%%%%%%%%%%%%%%%%%%%%%%%%%%%%%%%%%%%%%%
%%%%%%%%%%%%%%%%%%%%%%%%%%%%%%%%%%%%%%%%%%%
\section{Type IIB supergravity and \texorpdfstring{$AdS_6 \times S^2 \times \Sigma$}{AdS6xS2xSigma} 
Ansatz}
\label{sec:2}
\setcounter{equation}{0}
%%%%%%%%%%%%%%%%%%%%%%%%%%%%%%%%%%%%%%%%%%%
%%%%%%%%%%%%%%%%%%%%%%%%%%%%%%%%%%%%%%%%%%%

In this section, we provide a brief review of the Type IIB supergravity fields, Bianchi identities, field equations and BPS equations, and their $SU(1,1)$ duality symmetry, and go on to construct the general Ansatz for the bosonic fields of the solutions we seek to construct. As laid out in the introduction, the bosonic symmetries of the Ansatz are completely determined by the superconformal algebra.

\subsection{Type IIB supergravity review}

The bosonic fields of Type IIB supergravity \cite{Schwarz:1983qr,Howe:1983sra} consist of  the metric $g_{MN}$, a  one-form $P$ and gauge connection $Q$ representing the axion-dilaton field strengths, a complex three-form field strength $G$, and a self-dual five-form $F_5$ field strength. The  fields satisfy the following Bianchi identities,
\bea
0 &=& dP-2i Q\wedge P
\no \\
0 &=& d Q + i P\wedge \bar P
\no \\
0 &=& d G - i Q\wedge G +  P\wedge \bar G
\no \\
0 &=& 8 \, d F_{(5)} -  i G \wedge \bar G
\label{bianchi1}
\eea
The field strength $F_{(5)}$ is required to be self-dual,
\bea
\label{SDeq}
 F_{(5) } = * F_{(5) }
\eea
The field equations are given by,
\bea
0 & = &
\nabla ^M P_M - 2i Q^M P_M
+ {1\over 24} G_{MNP }G^{MNP}
\no \\
0 & = &
\nabla ^P  G_{MNP } -i Q^P G_{MNP}
- P^P \bar G_{MNP }
+ {2\over 3} i F_{(5)MNPQR }G^{PQR}
\no \\
0 & = & R_{MN }
- P_M  \bar P_N  - \bar P_M  P_N
- {1\over 6} (F_{(5)}^2)_{MN }
\no \\ && \hskip .5in
- {1\over 8} (G_M {} ^{PQ } \bar G_{N PQ }
+ {\bar G_M} {} ^{ PQ } G_{N PQ })
+{1\over 48 } g_{MN } G^{PQR } \bar G_{PQR }
\eea
The fermionic fields are the dilatino $\lambda$ and the gravitino $\psi_M$, both of which are complex Weyl spinors with opposite 10-dimensional chiralities, given by $\Gamma_{11} \lambda =\lambda$, and $\Gamma_{11} \psi_M  =-\psi_M$. The supersymmetry variations of the fermions  are,
\bea
\label{susy1}
\delta\lambda
&=& 
i (\G \cdot P) \cB^{-1} \ep^*
-{i\over 24} (\G \cdot G) \ep
\no \\
\delta \psi_M
&=& D _M  \ep
+ {i\over 480}(\G \cdot F_{(5)})  \Gamma_M  \ep
-{1\over 96}\left ( \Gamma_M (\G \cdot G)
+ 2 (\G \cdot G) \G_M \right ) \cB^{-1} \ep^*
\qquad
\eea
where ${\cal B}$ is the charge conjugation matrix of the Dirac-Clifford algebra.\footnote{Our convention for the signature of the 10-dimensional space-time metric  is $\eta = {\rm diag} (- + \cdots +)$; the Dirac-Clifford algebra is defined by the relations $\{ \Gamma ^M , \Gamma ^N \} = 2 \eta ^{MN} I_{32}$; and the charge conjugation matrix $\cB$ is defined by the relations $\cB \cB^*=I$ and $\cB \Gamma ^M \cB^{-1} = (\Gamma ^M)^*$.  We will use the convention that repeated indices are to be summed; complex conjugation will be denoted by {\sl bar} for functions and by {\sl star} for spinors; and we will use the notation $\G \cdot T \equiv \G^{M_1 \cdots M_p} T_{M_1 \cdots M_p}$ for the contraction of any antisymmetric tensor field $T$ of rank $p$ and the $\G$-matrix of the same rank.} The BPS equations are obtained by setting $\delta \lambda = \delta \psi _M=0$.

\sm

The Bianchi identities (\ref{bianchi1})  for the field strengths $P,Q,G,F_{(5)}$ can be solved  in terms of a complex scalar $B$;  a complex 2-form potential $C_{(2)}$, and a real 4-form  potential $C_{(4)}$. The fields $P$ and $Q$ are expressed  as follows,
\bea
\label{sugra1}
P  = f^2 d B \hskip 0.45in & \hskip 1in &  f^2=(1-|B|^2)^{-1}
\no \\
Q  =  f^2 \Im( B d  \bar B) &&
\eea
while the fields $G$ and $F_{(5)}$ are conveniently expressed in terms of $C_{(2)}$ and $C_{(4)}$ with the help of the complex field strength $F_{(3)} = d C_{(2)}$, 
\bea
\label{GF5}
G & = &  f \left ( F_{(3)} - B \bar F_{(3)} \right )
\no \\
F_{(5)} & = & dC_{(4)} + { i \over 16} \left ( C_{(2)} \wedge \bar F_{(3)}
- \bar C_{(2)} \wedge  F_{(3)} \right )
\eea
The scalar field $B$ is related to the complex scalar $\tau$ and the axion $\chi$, and dilaton $\phi$  by,
\bea
\label{Btau}
B = {1 +i \tau \over 1 - i \tau } \hskip 1in \tau =  \chi + i e^{- 2\phi}
\eea
The expectation value of  $e^{2 \phi}$ is related to the string coupling constant.

\subsection{\texorpdfstring{$SU(1,1)$}{SU(1,1)} duality symmetry}

Type IIB supergravity is invariant under $SU(1,1) \sim SL(2,\RR)$ symmetry. This symmetry leaves the Einstein frame metric $g_{\mu \nu}$ as well as the 4-form $C_{(4)}$ invariant, acts on the field $B$ by M\"obius transformations, and acts on the 2-form $C_{(2)}$ and its complex conjugate $\bar C_{(2)}$ by a linear transformation,
\bea
\label{stransf}
B           & \to & B^s = {u B  + v \over \bar v B + \bar u}
\no \\
C_{(2)} & \to & C_{(2)}^s = u C_{(2)} + v \bar C_{(2)}
\eea
with $u,v \in \CC$ and $|u|^2 - |v|^2=1$. In this non-linear realization of $SU(1,1)$ on $B$, the field $B$ takes values in the coset  $SU(1,1) /U(1)_q$, and the fermions $\lambda$ and $\psi_\mu$ transform linearly under the isotropy gauge group $U(1)_q$ with composite gauge field $Q$. The transformation rules for the  field strengths are \cite{Schwarz:1983qr},
\bea
\label{su11a}
P & \to & P^s = e^{2 i \theta} P
\no \\
Q & \to & Q^s = Q + d \theta
\no \\
G & \to & G^s = e^{i \theta} G
\eea
where the phase $\theta$ is defined by,
\bea
\label{su11b}
e^{i \theta} = \left ( {v \bar B + u \over \bar v B + \bar u} \right )^\half
\eea
The $SU(1,1)=SL(2,\RR)$ symmetry will serve as a useful guide to organize the holomorphic data in our local solution. As is well-known, the invariance of Type IIB supergravity under the continuous group $SL(2,\RR)$ is reduced in Type IIB string theory to invariance under the discrete  $SL(2,\ZZ)$ S-duality symmetry,  due to the charge quantization of non-perturbative one-branes, five-branes and D-instantons. In the construction of supergravity solutions, we will always allow for the continuous symmetry.

\subsection{\texorpdfstring{The $AdS_6 \times S^2 \times \Sigma$}{AdS6xS2xSigma} Ansatz}\label{sec:2-3}

We seek a general Ansatz in Type IIB supergravity with the following symmetry group,
\bea
SO(2,5) \times SO(3)
\eea
The factor $SO(2,5)$ requires the geometry to contain  $AdS_6$, while the factor $SO(3)$ requires $S^2$, so that our space-time is given by,
\bea
AdS_6 \times S^2 \times \Sigma
\eea
Here $\Sigma$ stands for the remaining two-dimensional space over which the product $AdS_6 \times S^2$ is warped. In order for the above space to be a Type IIB supergravity geometry,  $\Sigma$ must carry an orientation as well as a Riemannian metric, and is therefore a  Riemann surface, possibly with boundary. This $SO(2,5)\times SO(3)$-invariant Ansatz for the metric can be written as,
\bea
ds^2 = f_6^2 \, ds^2 _{AdS_6} + f_2 ^2 \, ds^2 _{S^2} + ds^2 _\Sigma
\eea
where $f_6,f_2$ and $ds^2 _\Sigma$ are  functions of $\Sigma$ only. We introduce an orthonormal frame,
\bea
\label{frame1}
e^m & = & f_6 \, \hat e^m \hskip 1in m=0,1,2,3,4,5
\no \\
e^{i} \, & = & f_2 \, \hat e^{i}  \hskip 1.1in i =6,7
\no \\
e^a &  & \hskip 1.4in a=8,9
\eea
where $\hat e^m$ and $\hat e^{i}$ respectively refer to the orthonormal  frames for the spaces $AdS_6$ and  $S^2$ with unit radius, and $e^a$ is an orthonormal frame for the metric on  $\Sigma $. In particular, we have,
\bea
ds^2 _{AdS_6} & = & \eta _{mn}^{(6)}  \, \hat e^m \otimes \hat e^n \hskip 1in \eta^{(6)} = {\rm diag} (-+++++)
\no \\
ds^2 _{S^2} & = & \delta _{i j} \, \hat e^{i} \otimes \hat e^{j}
\no \\
ds^2 _\Sigma & = & \delta _{ab} \, e^a \otimes e^b
\label{frame2}
\eea
By $SO(2,5)\times SO(3)$-invariance, the fields $P,Q,G$ and $F_{(5)}$ are given as follows,
\bea
\label{PQdef}
P =  p_a \, e^a & \hskip 1in & \hskip 0.15in G = g_a \, e^a \wedge e^{67}
\no \\
Q = q_a \, e^a && F_{(5)}=0
\eea
where $e^{67} = e^6 \wedge e^7$.
The components $p_a, q_a$, and $g_a$ are complex. Note that the Bianchi identity for the five-form field (\ref{bianchi1}) is automatically satisfied with this Ansatz.

%%%%%%%%%%%%%%%%%%%%%%%%%%%%%%%%%%%%%%%%%%%
%%%%%%%%%%%%%%%%%%%%%%%%%%%%%%%%%%%%%%%%%%%
\section{Reducing the BPS equations}
\setcounter{equation}{0}
\label{sec:3}
%%%%%%%%%%%%%%%%%%%%%%%%%%%%%%%%%%%%%%%%%%%
%%%%%%%%%%%%%%%%%%%%%%%%%%%%%%%%%%%%%%%%%%%

The residual supersymmetries, if any, of a configuration of purely bosonic Type IIB supergravity fields are governed by the BPS equations of (\ref{susy1}). Our interest is in purely bosonic field configurations which preserve 16 independent supersymmetries given by the $AdS_6 \times S^2 \times \Sigma$ Ansatz of the preceding section. It will turn out that any such configuration automatically solves the Bianchi and field equations, and thus automatically provides a half-BPS solution to Type IIB supergravity.

\sm

In this section, we will reduce the BPS equations to the $AdS_6 \times S^2 \times \Sigma$ Ansatz by decomposing the supersymmetry parameter $\ep$ of (\ref{susy1})  onto the Killing spinors of $AdS_6 \times S^2$. We will expose the residual symmetries of the reduced BPS equations, and solve those reduced equations which are purely algebraic in the supersymmetry spinor components. This will produce simple algebraic expressions for the metric factors $f_2, f_6$ in terms of the spinors. The remaining reduced BPS equations will then gradually be solved for the remaining bosonic fields as well as for the residual supersymmetries in subsequent sections. The strategy employed here is very similar to the one used in \cite{D'Hoker:2007xy} and so our discussion will closely follow that work.

\subsection{Killing spinors}

The Killing spinor equations on $AdS_6 \times S^2$ are,\footnote{The decomposition of the 10-dimensional Dirac-Clifford matrices under the reduction to the $AdS_6 \times S^2 \times \Sigma$ Ansatz, and the details of the Killing spinor equations, are relegated to appendices \ref{appA} and \ref{appB} respectively.}
\bea
\label{3a1}
\left ( \hat \nabla _m - { 1 \over 2} \eta _1 \, \gamma _m \otimes I_2 \right ) \chi ^{\eta _1, \eta _2} _\alpha& = & 0
\no \\
\left ( \hat \nabla _i - { i \over 2} \eta _2 \, I_8 \otimes \gamma _i \right ) \chi ^{\eta _1, \eta _2} _\alpha & = & 0
\eea
where $\hat \nabla _m$ and $\hat \nabla_i$ stand for the covariant spinor derivatives respectively on the spaces $AdS_6$ and $S^2$ with unit radius. Recall that $m$, $i$, and $a$ are all {\sl frame indices}. The spinors $\chi ^{\eta _1, \eta _2}_\alpha $ are 16-dimensional, and the parameters $\eta _1$ and $\eta_2$ can take the values $\pm 1$.  The solutions to these equation are 4-fold degenerate for each value of $\eta _1, \eta _2$,  and this degeneracy  will be labeled by the index $\alpha =1,2,3,4$. The chirality matrices act as follows,
\bea
\label{3a2}
\left ( \gamma _{(1)} \otimes I_2 \right ) \, \chi ^{\eta _1, \eta _2}_\alpha  & = & \chi ^{- \eta _1, \eta _2}_\alpha
\no \\
\left ( I_6 \otimes \gamma _{(2)} \right ) \, \chi ^{\eta _1, \eta _2} _\alpha & = & \chi ^{ \eta _1, - \eta _2}_\alpha
\eea
The way these equations should be understood is as follows. We begin with $\eta _1 = \eta _2=+$, and pick a basis $\chi _\alpha ^{++}$ for the four-dimensional vector space of spinors for fixed $\eta _1, \eta _2$
such that the action of $\gamma _{(1)}$ and $\gamma _{(2)}$ are diagonal. Then, we can simply define the basis for $\chi _\alpha ^{\eta _1 , \eta _2}$ for the remaining three values of $\eta_1, \eta _2$ by the action of the chirality matrices above.

\sm

Since $\ep^*$ appears in the fermion variations of (\ref{susy1}), we also need to understand how to express the complex conjugate spinor in this basis. If $\chi ^{\eta _1, \eta _2} _\alpha$ satisfies (\ref{3a1}), then by complex conjugating the entire first equation and using $(\gamma _m)^* = B_{(1)} \gamma _m B_{(1)}^{-1}$, we conclude that $B_{(1)} ^{-1} \otimes I_2 (\chi^{\eta _1 , \eta _2}_\alpha)^*$ satisfies the same equation, with the same values of $\eta_1, \eta _2$. Proceeding analogously for the second equation, we conclude that $I_6 \otimes B_{(2)} ^{-1} (\chi^{\eta _1 , \eta _2}_\alpha )^*$ also satisfies the same equation, with the same values of $\eta_1, \eta _2$. As a result, we must have the following linear relation,
\bea
( B_{(1)} ^{-1} \otimes B_{(2)}^{-1} ) (\chi^{\eta _1, \eta _2} _\alpha )^*  & =  &\sum _{\beta =1} ^4 M_{\alpha \beta}^{\eta _1, \eta _2} \, \chi ^{\eta _1, \eta _2} _\beta
\eea
for some matrix $M_{\alpha \beta}$ for each pair $\eta _1, \eta _2$.

\sm

We will now show that one may choose a basis for the Killing spinors $\chi ^{\eta _1, \eta _2}_\alpha $ in which $M=I$. Iterating the complex conjugation, we conclude that $(M^{\eta _1, \eta _2})^* (M^{\eta _1, \eta _2}) =I$, for all values of $\eta _1, \eta _2$. Specializing first to $\eta _1=\eta _2=+$, we have a single matrix $M^{+,+}$ satisfying $(M^{+,+})^* (M^{+,+})=I$. Now every such matrix may be rotated to the identity by a general linear complex-valued $4 \times 4$ matrix $U$, using the relation  $M^{+,+} = (U^*)^{-1}\, U$.
An easy way to construct $U$ is as follows. An arbitrary invertible complex  matrix $M$ in $GL(4,\CC)$ may be written as an exponential, $M = \exp ( i H)$ of a complex matrix $H$. Given $M$, the matrix $H$ is not unique. The condition $M^* M=I$ requires $H$ to be real-valued. Thus, we choose the solution $U= \exp \left ( { i \over 2} H \right )$ to the equation $M^{+,+} = (U^*)^{-1} \, U$, and the relation for $\eta _1=\eta _2=+$ may be diagonalized as follows,
\bea
( B_{(1)} ^{-1} \otimes B_{(2)}^{-1} ) (\chi^{+,+} _\alpha )^*   & = &   \chi ^{+,+} _\alpha
\eea
for $\alpha =1,2,3,4$. For the other values of $\eta_1, \eta _2$ we use (\ref{3a2}) to
express $\chi ^{\eta_1, \eta _2} _\alpha$ in terms of $\chi^{+,+} _\alpha$,
\bea
\chi ^{+,+} _\alpha
& = & (\gamma _{(1)} \otimes I_2) \, \chi ^{-,+}_\alpha
\no \\
& = & (I_6 \otimes \gamma _{(2)} ) \, \chi ^{+,-}_\alpha
\no \\
& = & (\gamma _{(1)} \otimes \gamma _{(2)} ) \, \chi ^{-,-}_\alpha
\eea
Using the fact that $\gamma _{(1)}$ commutes with $B_{(1)}$, while $\gamma _{(2)}$ anti-commutes with $B_{(2)}$, we find,
\bea
( B_{(1)} ^{-1} \otimes B_{(2)}^{-1} ) (\chi^{\eta _1, \eta _2} _\alpha )^*  & = &
\eta _2 \, \chi ^{\eta _1, \eta _2}_\alpha
\eea
for all values of $\eta _1, \eta _2, \alpha$.
Since this decomposition is now canonical in terms of the degeneracy index $\alpha$, we will no longer indicate it explicitly.

\subsection{Decomposing onto Killing spinors}

An arbitrary 32-component complex spinor $\ep$ may be decomposed onto the above Killing spinors
as follows,
\bea
\ep = \sum _{\eta _1, \eta _2 = \pm} \chi ^{\eta _1, \eta _2} \otimes \zeta _{\eta _1, \eta _2}
\eea
where $\zeta _{\eta _1, \eta_2}$ is a complex 2-component spinor for each $\eta _1, \eta _2$,
and the 4-fold degeneracy has not been indicated explicitly. As a supersymmetry generator in Type IIB, the spinor $\ep$ must be of definite chirality $\Gamma ^{11} \ep = - \ep$, which places the following chirality requirements on $\zeta$,
\bea
\gamma _{(3)} \zeta _{-\eta _1, -\eta _2} = - \zeta _{\eta _1, \eta _2}
\eea
The charge conjugate spinor is given by,
\bea
\cB^{-1} \ep ^* = i B_{(1)}^{-1} \otimes \gamma _{(2)} B_{(2)} ^{-1}  \otimes B_{(3)}^{-1}
\sum _{\eta _1, \eta _2} (\chi ^{\eta _1, \eta _2})^* \otimes \zeta _{\eta _1, \eta _2}^*
\eea
Since $\gamma _{(2)}$ anticommutes with $B_{(2)}$, we obtain,
\bea
\cB^{-1} \ep ^* & = &  i   I_8 \otimes \gamma _{(2)} \otimes I_2
\sum _{\eta _1, \eta _2} \left ( B_{(1)}^{-1} \otimes B_{(2)} ^{-1} (\chi ^{\eta _1, \eta _2})^* \right ) \otimes \left ( B_{(3)}^{-1}  \zeta _{\eta _1, \eta _2}^* \right )
\eea
It will be convenient to denote the result as follows,
\bea
\cB^{-1} \ep ^* & = &
\sum _{\eta _1, \eta _2} \chi ^{\eta _1,  \eta _2}  \otimes \star \zeta _{\eta_1, \eta_2}
\hskip 1in
\star \zeta _{\eta _1, \eta_2} =  - i \eta _2 \sigma ^2  \zeta _{\eta _1, - \eta _2}^*
\eea
As in \cite{D'Hoker:2007xy} we will use the $\tau$ matrix notation introduced originally in \cite{Gomis:2006cu} in order to compactly express the action of the various $\gamma$ matrices on $\zeta$. Defining $\tau^{(ij)} = \tau^i \otimes \tau^j$ with $i,j=0,1,2,3$, $\tau^0$ the identity matrix and $\tau^i$ with $i=1,2,3$
the standard Pauli matrices, we can write,
\bea
(\tau^{(ij)} \zeta)_{\eta_1,\eta_2} \equiv \sum_{\eta_1',\eta_2'} (\tau^i)_{\eta_1 \eta_1'} (\tau^j)_{\eta_2 \eta_2'} \zeta_{\eta_1' \eta_2'}
\eea

\subsection{Symmetries of the reduced BPS equations}

Using the decomposition of $\ep$ into Killing spinors and the streamlined notation of the $\tau$ matrices we can finally write down the BPS equations in a reduced form. The reduced dilatino equation is,
\bea
0 = - 4 p_a \gamma ^a \sigma ^2 \zeta ^* +  g_a \tau^{(03)}  \gamma ^a \zeta
\eea
while the reduced gravitino equations take the following form,
\bea
\label{redmia}
(m) &\qquad&
0 = - {i \over  2f_6} \tau^{(21)}  \zeta
+ {D_a f_6 \over  2f_6}   \gamma^{a} \zeta
- {1 \over 16}  g_a \tau^{(03)}  \gamma^{a}  \sigma^2 \zeta^*
\\
(i) &\qquad&
0 = {1 \over  2 f_2} \tau^{(02)}  \zeta
+ {D_a f_2 \over  2f_2}   \gamma^{a} \zeta
+ {3  \over 16}  g_a \tau^{(03)} \gamma^{a}  \sigma^2 \zeta^*
\no\\
(a) &\qquad&
0 = \bigg( D_a  + {i \over 2} \hat \omega _a   \sigma^{3} - { i \over 2} q_a \bigg) \zeta
+ {3  \over 16} g_a \tau^{(03)}   \sigma^2 \zeta^*
- {1 \over 16}  g_b \tau^{(03)}  \gamma_a {}^b  \sigma^2 \zeta^*
\no
\eea
The derivative $D_a$  is defined with respect to the frame $e^a$, so that the total differential $d$ takes the form $d=e^a D_a$, while the $U(1)$-connection with respect to frame indices is $\hat \omega _a$.

\sm

Before we move on to solving these equations, let us briefly look at their symmetries. The axion/dilaton field $B$ transforms non-linearly under $SU(1,1)$ of Type IIB supergravity and takes values in the coset $SU(1,1)/U(1)_q$. Global $SU(1,1)$ transformations on the fields are accompanied by local $U(1)_q$ gauge transformations, given in (\ref{su11a}), and which induce the following symmetry transformations on the fields of the reduced BPS equations,
\bea
U(1)_q & \hskip 1in & \zeta \, \to \, e^{i \theta /2} \, \zeta
\no \\
&& q_a \, \to \, q_a + D_a \theta
\no \\
&& p_a \, \to \, e^{2 i \theta } \, p_a
\no \\
&& g_a \, \to \, e^{i \theta } \, g_a
\eea
The reduced BPS equations are also invariant under the following discrete symmetries,
\bea
\zeta & \to & \cI \, \zeta = - \tau ^{(11)} \, \sigma ^3 \, \zeta
\no \\
\zeta & \to & \cJ \, \zeta = \tau ^{(32)} \, \zeta
\eea
which leave all the bosonic fields invariant.
Both $\cI$ and $\cJ$ commute with $U(1)_q$.
Finally, complex conjugation is naturally combined with $U(1)_q$, and we have,
\bea
\zeta & \to & \cK \, \zeta \, = e^{i \theta } \, \tau ^{(22)} \, \sigma ^2 \, \zeta ^*
\no \\
q_a & \to & \cK \, q_a = - q_a + 2 D_a \theta
\no \\
p_a & \to & \cK \, p_a = e^{4 i \theta } \, \bar p_a
\no \\
g_a & \to & \cK \, g_a = - e^{ 2 i \theta } \, \bar g_a
\eea
The chirality requirement of Type IIB restricts the spinor $\zeta$ to the subspace,
\bea
\label{chireq}
\cI \, \zeta = - \tau^{(11)} \, \sigma ^3 \, \zeta = \zeta
\eea
In the next subsection, we will investigate the restrictions to the eigen-spaces of $\cJ$ and $\cK$ imposed by the reduced BPS equations. The symmetries $\cI, \cJ, \cK$ commute with one another, so that we may diagonalize them simultaneously, and restrict to any one of their common subspaces.

\subsection{Restricting to a single subspace of \texorpdfstring{$\cJ$}{J}}

We will assume that $p_a$ does  not vanish  identically. We now use the dilatino equation to derive a first set of bilinear relations. Multiply the dilatino equation to the left by $\zeta ^t T \sigma ^2$ and choose the
$\tau$-matrix $T$ so that the flux part vanishes,
\bea
g_a \zeta ^t T  \tau ^{(03)} \sigma ^2 \gamma^a \zeta =0
\eea
Since $\sigma ^2 \gamma ^a$ is symmetric for $a =8,9$, this condition will require $T$ to satisfy the condition that the product $T \tau^{(03)}$ is anti-symmetric, which has the following solutions,
\bea
T \in \cT= \{ \tau ^{(01)}, \,  \tau ^{(11)}, \,  \tau ^{(31)}, \,  \tau ^{(20)}, \,  \tau ^{(22)}, \,  \tau ^{(23)} \}
\eea
The equation implied on $p_a$ is then given by the complex conjugate of,
\bea
\bar p_a \zeta ^\dagger T  \gamma ^a  \zeta =0
\eea
When $p_a \not=0$, we can draw from this equation only an orthogonality relation. To obtain a full vanishing condition, we make further use of the chirality condition, and obtain,
\bea
\bar p_a \zeta ^\dagger T  \tau ^{(11)} \gamma ^a  \sigma ^3 \zeta =0
\eea
We obtain such a relation when both $T $ and $T\tau^{(11)}$ belong to $\cT$, which is the case for
only a single pair, namely $T=\tau ^{(20)} , \tau ^{(31)}$. As a result, we have the equivalent relations,
\bea
\label{3d4}
\zeta ^\dagger \tau ^{(20)} \gamma ^a \zeta = \zeta ^\dagger \tau ^{(31)} \gamma ^a \zeta=0
\eea

\sm

Next, we analyze the gravitino equations. We multiply equations $(m)$ and $(i)$ of (\ref{redmia}) on the left by $\zeta ^\dagger T \sigma ^p$ for $p=0,3$, and obtain cancellation of the last term when $T \tau ^{(03)}$ is antisymmetric (which is the same condition we had for the dilatino equation),
\bea
0 & = & - {i \over  2f_6} \zeta ^\dagger T \tau^{(21)} \sigma ^p \zeta
+ {D_a f_6 \over  2f_6} \zeta ^\dagger T  \sigma ^p \gamma^{a} \zeta
\no \\
0 & = & {1 \over  2 f_2} \zeta ^\dagger T \tau^{(02)} \sigma ^p  \zeta
+ {D_a f_2 \over  2f_2} \zeta ^\dagger T \sigma ^p \gamma^{a} \zeta
\eea
In view of (\ref{3d4}), the second term will cancel when $T= \tau ^{(20)}$ and $T = \tau ^{(31)}$, so that we obtain the following relations from the remaining cancellation of the first term,
\bea
\zeta ^\dagger \tau ^{(01)} \sigma ^p \zeta & = & 0  \hskip 1in p=0,3
\no \\
\zeta ^\dagger \tau ^{(22)} \sigma ^p \zeta & = & 0
\eea
and their chiral conjugates, obtained by using the chirality condition $\tau^{(11)} \sigma ^3 \zeta = - \zeta$,
\bea
\zeta ^\dagger \tau ^{(10)} \sigma ^p \zeta & = & 0 \hskip 1in p=0,3
\no \\
\zeta ^\dagger \tau ^{(33)} \sigma ^p \zeta & = & 0
\eea
Next, we use the general result of \cite{D'Hoker:2007xy} that the bilinear equation $\zeta ^\dagger M \zeta=0$  is  solved by projecting $\zeta$ onto a subspace with the help of a projection matrix $P$ that anti-commutes with $M$. This result was established for the case of 2-dimensional $M$, which is in fact the case also here by reduction. Thus, we must find a projector $P$ which commutes with $\cI$ and with the following properties,
\bea
[ P, \tau ^{(11)} \sigma ^3 ] = \{ P, \tau^{(01)} \sigma ^p \} = \{ P, \tau ^{(22)} \sigma ^p \} =0
\eea
The solutions to these conditions are $\tau ^{(32)}$ and $\tau^{(23)}$, possibly multiplied by a factor of $\sigma ^3$. These four  possibilities are pairwise equivalent under the chirality relation. Now the projector $P=\tau ^{(32)}$ precisely corresponds to the symmetry $\cJ$, so imposing a restriction on the spinor space by this operator is the only consistent restriction. Therefore, we will impose,
\bea
\label{projector}
\tau ^{(32)} \zeta = \nu \zeta \hskip 1in \nu = \pm 1
\eea
which solves all the above bilinear relations. One must pick one value of $\nu$ or the other in the projection.

\sm

Imposing the chirality relation (\ref{chireq}), $\tau ^{(11)} \, \sigma ^3 \, \zeta  =  - \zeta$, as well as the projector
(\ref{projector}) we just derived,
we may solve for the relations between the components of $\zeta$. Denoting the components by $\zeta _{abc}$, which take values $\pm$, the $a,b$ components labels the $\tau$-matrix basis, while $c$ labels the chirality basis in which $\sigma ^3$ is diagonal. We then have two independent complex-valued components which we denote by $\alpha$ and $\beta$, and which are defined as follows,
\bea
\bar \alpha & = & \zeta _{+++} \, = \, - \zeta _{--+} \, = \, - i \nu \zeta _{+-+} \, = \, + i \nu \zeta _{- ++}
\no \\
\beta & = & \zeta _{---} \, = \, + \zeta _{++-} \, = \, - i \nu \zeta _{-+-} \, = \, - i \nu \zeta _{+--}
\label{spinorsolution}
\eea
To reduce the equations to a basis of complex frame indices $z, \bar z$, we will use the following conventions,
\bea
e^z = \half \left ( e^8 + i e^9 \right ) & \hskip 1in &
\gamma ^z = \half \left ( \gamma ^8 + i \gamma ^9 \right )
= \begin{pmatrix} 0 & 1 \cr 0 & 0 \end{pmatrix}
\no \\
e^{\bar z} = \half \left ( e^8 - i e^9 \right ) & \hskip 1in &
\gamma ^{\bar z} = \half \left ( \gamma ^8 - i \gamma ^9 \right )
= \begin{pmatrix} 0 & 0 \cr 1 & 0 \end{pmatrix}
\eea
In particular, we have,
\bea
\delta _{z \bar z} =2 \hskip 1in \delta ^{z \bar z} = \half
\eea
It will also be convenient to have the following results of $\gamma _a {}^b \, \sigma ^2$,
\bea
\gamma _z {} ^z \, \sigma ^2 = - \gamma _{\bar z} {} ^{\bar z} \, \sigma ^2 = i \sigma ^1
\eea

\subsection{The reduced BPS equations in component form}

Using the solution (\ref{spinorsolution}) we found for the projection condition on the preserved supersymmetries, the reduced dilatino equations become,
\bea
\label{dilBPS}
4 i p_z \, \alpha - g_z \, \beta & = & 0
\no \\
4 i p_{\bar z} \, \bar \beta  + g_{\bar z} \, \bar \alpha & = & 0
\eea
The algebraic gravitino equations are,
\bea
\label{gravBPS1}
{ 1 \over 2 f_6} \, \bar \alpha +{ D_z f_6 \over 2 f_6} \, \beta -{ i \over 16} g_z \, \alpha   & = & 0
\no \\
- { 1 \over 2 f_6} \, \beta +{ D_{\bar z} f_6 \over 2 f_6} \, \bar \alpha  +{ i \over 16} g_{\bar z} \, \bar \beta  & = & 0
\no \\
{ \nu \over 2 f_2} \, \bar \alpha +{ D_z f_2 \over 2 f_2} \, \beta + {3 i \over 16} g_z \, \alpha  & = & 0
\no \\
{ \nu \over 2 f_2} \, \beta +{ D_{\bar z} f_2 \over 2 f_2} \, \bar \alpha  -{ 3i \over 16} g_{\bar z} \, \bar \beta   & = & 0
\eea
Finally, the component decomposition of the differential equations is as follows,
\bea
\label{gravBPS2}
\left ( D_z + { i \over 2} \hat \omega _z - { i \over 2} q_z \right ) \bar \alpha - { i \over 4} g_z \bar \beta  & = & 0
\no \\
\left ( D_z - { i \over 2} \hat \omega _z - { i \over 2} q_z \right ) \beta + { i \over 8} g_z \alpha  & = & 0
\no \\
\left ( D_{\bar z}  + { i \over 2} \hat \omega _{\bar z} - { i \over 2} q_{\bar z}  \right ) \bar \alpha
- { i \over 8} g_{\bar z} \bar \beta  & = & 0
\no \\
\left ( D_{\bar z}  - { i \over 2} \hat \omega _{\bar z} - { i \over 2} q_{\bar z}  \right ) \beta
+ { i \over 4} g_{\bar z} \alpha  & = & 0
\eea
In addition, we have the complex conjugate equations to all of the equations above.
Note that since $G$ and $P$ are complex-valued, we have in general $(g_z)^* \not= g_{\bar z}$ and $(p_z)^* \not= p_{\bar z}$.

\subsection{Determining the radii \texorpdfstring{$f_6, f_2$}{f6, f2} in terms of the spinors}

We begin with the two equations that involve $D_z f_6$,
\bea
{ 1 \over 2 f_6} \, \bar \alpha +{ D_z f_6 \over 2 f_6} \, \beta -{ i \over 16} \, g_z \, \alpha   & = & 0
\no \\
- { 1 \over 2 f_6} \, \bar \beta  +{ D_z f_6 \over 2 f_6} \, \alpha  -{ i \over 16} (g_{\bar z} )^*  \, \beta  & = & 0
\eea
To eliminate the contributions from the first term in each equation,  we add the first line times $\bar \beta$ to the second line times $\bar \alpha$, and we obtain,
\bea
\label{Df6}
{ D_z f_6 \over 2 f_6} (\alpha \bar \alpha  + \beta \bar \beta )
= { i \over 16} g_z \alpha \bar \beta  + { i \over 16} (g_{\bar z} )^* \bar \alpha \beta
\eea
Proceeding analogously for $D_z f_2$, we have,
\bea
{ \nu \over 2 f_2} \, \bar \alpha +{ D_z f_2 \over 2 f_2} \, \beta + { 3 i \over 16} \, g_z \, \alpha   & = & 0
\no \\
 { \nu \over 2 f_2} \, \bar \beta  +{ D_z f_2 \over 2 f_2} \, \alpha   + { 3i \over 16} (g_{\bar z} )^*  \, \beta  & = & 0
\eea
To eliminate the contributions from the first term in each equation, we add the first line times $- \bar \beta$ to the second line times $\bar \alpha$, and we obtain,
\bea
\label{Df2}
{ D_z f_2 \over 2 f_2} (\alpha \bar \alpha  - \beta \bar \beta )
= {3 i \over 16} g_z \alpha \bar  \beta  - { 3i \over 16} (g_{\bar z} )^* \bar \alpha \beta
\eea
These combinations suggest that we should evaluate the  covariant derivatives $D_z( \alpha \bar \alpha \pm \beta \bar \beta)$   out of  the differential equations (\ref{gravBPS2}) for $\alpha, \beta, \bar \alpha , \bar \beta $, and we find,
\bea
\label{Dab}
D_z (\alpha \bar \alpha  + \beta \bar \beta ) & = &
{ i \over 8} g_z \alpha \bar \beta  + { i \over 8} (g_{\bar z} )^* \bar \alpha \beta
\no \\
D_z (\alpha \bar \alpha  - \beta \bar \beta ) & = &
{ 3i \over 8} g_z \alpha \bar  \beta  - { 3 i \over 8} (g_{\bar z} )^* \bar \alpha \beta
\eea
We may now eliminate all flux dependences between (\ref{Df6}), (\ref{Df2}) and (\ref{Dab}), and we find,
\bea
\label{f2f7}
{ D_z f_6 \over  f_6} (\alpha \bar \alpha  + \beta \bar \beta )  & = & D_z (\alpha \bar \alpha  + \beta \bar \beta )
\no \\
{ D_z f_2 \over  f_2} (\alpha \bar \alpha  - \beta \bar \beta )  & = & D_z (\alpha \bar \alpha  - \beta \bar \beta )
\eea
Given that the arguments of the derivatives are real functions, we conclude,
\bea
\label{f2f6}
f_6 & = & c_6 \, \left ( \alpha \bar \alpha  + \beta \bar \beta  \right )
\no \\
f_2 & = & c_2 \, \left ( \alpha \bar \alpha  - \beta \bar \beta  \right )
\eea
for constant $c_2, c_6$. In obtaining the equation for $f_2$ from the last equation in (\ref{f2f7}), we have assumed that $|\alpha|^2-|\beta |^2 $ does not vanish identically.

\subsection{Solving the remaining algebraic gravitino equations}

To obtain the results of the previous subsection, we have taken only pairwise linear combinations of the algebraic gravitino equations. Here, we take the orthogonally conjugate pairwise linear combinations, multiplying the first equation by $\bar \alpha $ and the second by $- \beta$. The determinant of the two linear combinations is $ \alpha \bar \alpha  + \beta \bar \beta \not=0$, so that the four resulting bilinear equations are guaranteed to be equivalent to the original four algebraic gravitino equations. The terms in $D_z f_6$ and $D_z f_2$ cancel out, and we are left with,
\bea
{ 1 \over 2 c_6}  - { i \over 16} g_z \alpha^2 + { i \over 16} (g_{\bar z} )^* \beta ^2 & = & 0
\no \\
{ \nu \over 2 c_2}  + { 3 i \over 16} g_z \alpha ^2 - {3 i \over 16} (g_{\bar z} )^* \beta ^2 & = & 0
\eea
The last equation may be simplified with the help of the first and yields,
\bea
c_2 = - { \nu \over 3} \, c_6
\eea
Recall that $\nu$ can take the values $\nu=\pm 1$.

\subsection{Summary of remaining equations}

We may summarize the remaining equations as follows. The dilatino equations are the only ones involving $p_z$, and may be viewed as determining $p_z$,
\bea
\label{dil}
4 i p_z \, \alpha  - g_z \, \beta & = & 0
\no \\
4 i p_{\bar z} \, \bar \beta  + g_{\bar z} \, \bar \alpha & = & 0
\eea
Next, we have the radii in terms of the spinors,
\bea\label{f2f6-2}
f_6 & = & c_6 \, \left ( \alpha \bar \alpha  + \beta \bar \beta  \right )
\no \\
f_2 & = & - { \nu \over 3} \, c_6 \, \left ( \alpha \bar \alpha  - \beta \bar \beta  \right )
\eea
and the algebraic relation between the spinors and the fluxes,
\bea
\label{alg}
{ 1 \over 2 c_6}  - { i \over 16} g_z \alpha ^2 + { i \over 16} (g_{\bar z} )^* \beta ^2  =  0
\eea
and finally the differential equations,
\bea\label{diffspin}
\left ( D_z - { i \over 2} \hat \omega _z + { i \over 2} q_z \right ) \alpha + { i \over 8} (g_{\bar z})^*  \beta & = & 0
\no \\
\left ( D_z - { i \over 2} \hat \omega _z - { i \over 2} q_z \right ) \beta + { i \over 8} g_z \, \alpha  & = & 0
\no \\
\left ( D_z + { i \over 2} \hat \omega _z - { i \over 2} q_z \right ) \bar \alpha - { i \over 4} g_z \, \bar \beta  & = & 0
\no \\
\left ( D_z + { i \over 2} \hat \omega _z + { i \over 2} q_z \right ) \bar \beta - { i \over 4} (g_{\bar z})^*  \bar \alpha & = & 0
\eea
along with their complex conjugates.

%%%%%%%%%%%%%%%%%%%%%%%%%%%%%%%%%%%%%%%%%%%
%%%%%%%%%%%%%%%%%%%%%%%%%%%%%%%%%%%%%%%%%%%
\section{Local solutions to the BPS equations}
\setcounter{equation}{0}
\label{sec:4}
%%%%%%%%%%%%%%%%%%%%%%%%%%%%%%%%%%%%%%%%%%%
%%%%%%%%%%%%%%%%%%%%%%%%%%%%%%%%%%%%%%%%%%%

In the previous section the BPS equations were reduced and solved for the radii $f_6, f_2$. The remaining equations are all complex-valued and are organized as follows: we have one algebraic relation (\ref{alg}) and four differential equations (\ref{diffspin}) for the spinor components $\alpha, \beta$, the reduced flux fields $g_z, g^*_z$ and their complex conjugates. Two further algebraic equations give the axion-dilaton field $p_a$ in terms of the spinors and the fluxes. In this section we will solve completely for the local solutions to this system. Specifically, we will derive expressions for all supergravity fields  that satisfy the BPS equations, the Bianchi identities, and the supergravity field equations, in terms of two locally holomorphic functions $\cA_+,\cA_-$ on $\Sigma$. We will discuss the conditions on the  local solutions imposed by the proper Minkowski signature of the metric and the absence of singularities  in section  \ref{sec:5}.

\sm

The construction of the local solution is involved and proceeds in a number of steps which we will now outline, and carry out in this section in detail. First, we will eliminate the reduced flux fields $g_z$ and $g_{\bar z}$  in favor of the reduced axion-dilaton fields $p_z, p_{\bar z}$ and similarly for their complex conjugates. Second, we will use the expression for the reduced axion-dilaton fields $p_z, p_{\bar z}, q_z, q_{\bar z} $ in terms of $\rho$ and  $B$ to decouple and integrate one pair of the differential equations, and  obtain the spinor components $\bar \alpha, \bar \beta$ in terms of two  holomorphic one-forms $ \kappa_\pm$ as well as $B$ and  $\rho$. Third, we will eliminate the spinor components $\alpha, \beta$  in favor of $\kappa_\pm$, $\rho$, and $B$ as well, and thereby produce three nonlinear partial differential equations for the complex field $B$ and the real field $\rho$. Being non-linear, these equations are not easy to solve, and give rise to a situation reminiscent 
of \cite{D'Hoker:2007xy}. However, in a fourth step we will identify a sequence of two changes of variables which decouples these non-linear differential equations. In a final fifth step, we will solve all decoupled equations in terms of two holomorphic functions $\cA_\pm$ which are related to $\kappa_\pm$ by $\kappa_\pm =\partial_w \cA_\pm$.  For the reader who wishes to skip this entire derivation we have summarized the final result  in the introduction to section~\ref{sec:5-2}.

\subsection{Eliminating the reduced flux fields {\rm (step 1/5)}}

First, we eliminate the reduced flux fields $g_z, g_{\bar z}$ and their complex conjugates  in favor of $p_z, p_{\bar z} $ and their complex conjugates, using the dilatino BPS equations (\ref{dil}),
\bea
\label{5a1}
g_z = 4 i p_z \, { \alpha \over \beta}
\hskip 1in
(g_{\bar z} )^* = 4 i (p_{\bar z} )^* \, { \beta \over \alpha}
\eea
The  algebraic relation (\ref{alg}) becomes,
\bea\label{algb}
p_z \, { \alpha ^3 \over \beta} - (p_{\bar z})^* \, {\beta ^3 \over \alpha} + { 2 \over c_6}=0
\eea
The differential equations (\ref{diffspin}) take the following form,
\bea
\label{diffspinb}
\left ( D_z - { i \over 2} \hat \omega _z + { i \over 2} q_z \right ) \alpha
- \half  (p_{\bar z})^* \, { \beta ^2 \over \alpha} & = & 0
\no \\
\left ( D_z - { i \over 2} \hat \omega _z - { i \over 2} q_z \right ) \beta
- \half  p_z \, { \alpha^2 \over \beta}   & = & 0
\no \\
\left ( D_z + { i \over 2} \hat \omega _z - { i \over 2} q_z \right ) \bar \alpha
+ p_z  \, { \alpha \bar \beta \over \beta}  & = & 0
\no \\
\left ( D_z + { i \over 2} \hat \omega _z + { i \over 2} q_z \right ) \bar \beta
+ (p_{\bar z})^*  { \bar \alpha \beta \over \alpha} & = & 0
\eea
Equations (\ref{algb}) and (\ref{diffspinb}) are the remaining relations to be solved. The integrability conditions on the differential equations reproduce the Bianchi identities for the fields $P$ and~$Q$.

\subsection{Integrating the first pair of differential equations {\rm (step 2/5)}}

In the second step, we show that the first two equations of (\ref{diffspinb}) can be solved in terms of holomorphic functions. Multiplying the first equation of (\ref{diffspinb}) by $\alpha$ and the second equation of (\ref{diffspinb}) by $\beta $, we get equivalently,
\bea
\label{onetwoeqa}
\left ( D_z -  i  \hat \omega _z +  i  q_z \right ) \alpha^2 -   (p_{\bar z})^* \,  \beta ^2
& = & 0
\no \\
\left ( D_z -  i  \hat \omega _z -  i  q_z \right ) \beta ^2 -   p_z \,  \alpha^2    & = & 0
\eea
We switch to conformally flat complex coordinates $(w,\bar w)$ on $\Sigma$,
such that the metric reads $ds^2_\Sigma=4\rho^2 dw d\bar w$ and we have,
\begin{align}\no
 e^z &= \rho dw &  D_z&=\rho^{-1} \partial_w
 &
 \hat \omega_z&=i\rho^{-2}\partial_w \rho
 \\
 e^{\bar z}&=\rho d\bar w &
 D_{\bar z}&=\rho^{-1}\partial_{\bar w}
 &
 \hat \omega_{\bar z}&=-i\rho^{-2}\partial_{\bar w} \rho
 \label{eq:w-coords}
\end{align}
The extra factor of $\rho^{-1}$ in the derivatives $D_z$, $D_{\bar z}$ is due to $z$, $\bar z$ being frame indices.
We then express $p_z$ and $q_z$ in terms of the complex field $B$ using (\ref{sugra1}),
to recast (\ref{onetwoeqa}) as follows,
\bea\label{onetwob}
\p_w (\rho \alpha ^2) & = &
- \half f^2 \Big ( B \p_w \bar B - \bar B \p_w B \Big ) \rho \alpha ^2 + f^2 (\p_w \bar B) \rho \beta ^2
\no \\
\p_w (\rho \beta ^2) & = &
+ \half f^2 \Big ( B \p_w \bar B - \bar B \p_w B \Big ) \rho \beta ^2 + f^2 (\p_w B) \rho \alpha ^2
\eea
By taking suitable linear combinations it is straightforward to verify  that the following two equations are equivalent to (\ref{onetwob}),
\bea
\p_w \Big ( \ln  \{ \rho ( \alpha ^2 - \bar B \beta ^2)  \} + \ln f \Big ) & = & 0
\no \\
\p_w \Big ( \ln \{ \rho ( B \alpha ^2 - \beta ^2) \} + \ln f \Big ) & = & 0
\eea
These equations are solved in terms of two independent holomorphic 1-forms
$\kappa _\pm$, as follows,
\bea
\label{5b3}
\rho f \Big (  \alpha ^2 -  \bar B \beta ^2 \Big ) & = & \bar \kappa _+
\no \\
\rho f \Big ( \beta ^2 - B \alpha ^2  \Big ) & = & \bar \kappa _-
\eea
Inverting the relation (\ref{5b3}), we get expressions for the spinor components $\alpha,\beta$,
and their complex conjugates $\bar \alpha, \bar \beta$,
\bea
\label{5b4}
\rho \alpha ^2 = f (\bar \kappa _+ + \bar B \bar \kappa_-)
& \hskip 1in &
\rho \bar \alpha ^2 = f (\kappa _+ + B \kappa _-)
\no \\
\rho \beta ^2 = f ( B \bar \kappa _+ + \bar \kappa _-)
&&
\rho \bar \beta ^2 = f ( \bar B \kappa _+ + \kappa _-)
\eea
The right side of all four equations involves only the holomorphic data $\kappa _\pm$ and the $B$-field and their complex conjugates. It remains to solve for the $\rho$ and $B$-fields.

\subsection{Solving the second pair of differential equations {\rm (step 3/5)}}
\label{sec:4-2}

In the second step we express the third and fourth equation of (\ref{diffspinb}) in terms of $B,\rho$ and the local complex coordinates giving,
\bea
\Big ( \p _w - 2 (\p_w \ln \rho) - f^2 B \p_w \bar B \Big ) \left ( f \rho \bar \alpha ^2 \right )
+ 2 f^2 (\p_w B) f \rho { \alpha \bar \alpha \bar \beta \over \beta} & = & 0
\no \\
\Big ( \p _w - 2 (\p_w \ln \rho) - f^2 \bar B \p_w  B \Big ) \left ( f \rho \bar \beta ^2 \right )
+ 2 f^2 (\p_w \bar B) f \rho {  \bar \alpha \beta \bar \beta \over \alpha} & = & 0
\eea
Next, we compute the derivatives  of $\bar \alpha$ and $\bar \beta$ using (\ref{diffspinb}).
After taking suitable linear combinations, with coefficients $(\bar B, 1)$ and $(1, B)$ of the resulting equations, we find the  following equivalent system,
\bea
- (\p_w \bar B) f \rho \bar \alpha ^2 - 2 (\p_w \bar B) \rho f { \bar \alpha \beta \bar \beta \over \alpha}
+ { 2 \over f^2} (\p_w \ln \rho ) f \rho \bar \beta ^2 & = & \bar B \p_w \kappa _+ + \p_w \kappa _-
\no \\
- (\p_w  B) f \rho \bar \beta ^2 - 2 (\p_w B) \rho f { \alpha \bar \alpha  \bar \beta \over \beta}
+ { 2 \over f^2} (\p_w \ln \rho ) f \rho \bar \alpha ^2 & = &  \p_w \kappa _+ + B \p_w \kappa _-
\eea
One now eliminates the spinor components using (\ref{5b4})
in terms of $B$, $f$ and $\rho$, and we obtain, after some simplifications,
\bea
\label{5c5}
2 \p_w \ln \rho - f^2 (\p_w \bar B) \, { \kappa _+ + B \kappa _- \over \bar B \kappa _+ + \kappa _-}
- 2 f^2 (\p_w \bar B ) \, e^{+i \vartheta} & = & { \bar B \p_w \kappa _+ + \p_w \kappa _- \over \bar B \kappa _+ + \kappa _-}
\no \\
2 \p_w \ln \rho - f^2 (\p_w  B) \, { \bar B \kappa _+ +  \kappa _- \over  \kappa _+ + B \kappa _-}
- 2 f^2 (\p_w  B ) \, e^{-i \vartheta} & = & {  \p_w \kappa _+ + B \p_w \kappa _- \over  \kappa _+ + B \kappa _-}
\eea
Here we have used the following abbreviation for the phase angle $\tet$,
\bea
\label{phase1}
e^{2 i \tet} = \left ( { \kappa _+ + B \kappa _- \over \bar \kappa _+ + \bar B \bar \kappa _-}  \right )
\, \left ( { B \bar \kappa _+ + \bar \kappa _- \over \bar B \kappa _+ +  \kappa _-} \right )
\eea
Note that by subtracting the two equations (\ref{5c5}) the dependence on the metric factor $\rho$ can be eliminated  and one arrives at an equation for $B$ and $\bar B$ only in terms of $\kappa _\pm$,
\bea
\label{eqn:Beq}
&&
f^4 (\p_w B) (\bar B \kappa _+ + \kappa _-)^2 - f^4 (\p_w \bar B) ( \kappa _+ + B \kappa _-)^2
\no \\ &&
+ 2 f^4 ( \kappa _+ + B \kappa _-) ( \bar B \kappa _+ + \kappa _-)
\Big ( (\p_w B ) \, e^{- i \tet} - (\p_w \bar B) \, e^{+i \tet} \Big )
\no \\ && \hskip 0.3in
=
\kappa _+ \p_w \kappa _- - \kappa _- \p_w \kappa _+
\eea
The equations (\ref{5c5}) are of course supplemented by their complex conjugates. While  Eq.~(\ref{eqn:Beq}) seemingly depends on  both holomorphic one-forms $\kappa_+$ and  $\kappa_-$, in fact it depends only on their ratio. Assuming that $\kappa_-$ does not vanish identically, we  define the holomorphic or meromorphic  function $\lambda$ by,
\be
\lambda ={\kappa_+\over \kappa_-}
\ee
it is straightforward to show that  (\ref{eqn:Beq}) depends on $\lambda$ alone and takes the form,
\bea
\no
 f^4(\partial_w B)(\bar B\lambda+1)^2-f^4(\partial_w \bar B)(\lambda+B)^2\qquad\qquad\qquad&&\\
 +2f^4(\lambda+B)(\bar B\lambda+1)\left((\partial_w B)e^{-i\vartheta}-(\partial_w\bar B)e^{i\vartheta}\right)
 &=&-\partial_w\lambda
 \label{eqn:Beq-lambda}
\eea
where the phase of (\ref{phase1}) is now given by the expression, 
\bea
\label{thphase}
 e^{2i\vartheta}=\frac{(\lambda+B)(B\bar\lambda+1)}{(\bar\lambda+\bar B)(\bar B\lambda+1)}
\eea

\sm

The last equation we have to deal with is the algebraic relation (\ref{algb}).
As before, it  may be expressed in terms of $B,\rho$ and the local complex coordinates giving,
\bea
f^2 (\p_w B) \, {\rho \alpha ^3 \over \beta} - f^2 (\p_w \bar B) \, { \rho \beta ^3 \over \alpha} +{ 2 \rho ^2 \over c_6}=0
\eea
Eliminating $\rho \alpha ^2$ and $\rho \beta ^2$ using (\ref{5b4}), we obtain,
\bea
\label{5e1}
f^3 \, (\p_w B) \, { ( \bar \kappa _+ + \bar B \bar \kappa _-)^{3 \over 2} \over (B \bar \kappa _+ + \bar \kappa _-)^{1 \over 2}}
- f^3 \, (\p_w \bar B) \, { ( B \bar \kappa _+ +  \bar \kappa _-)^{3 \over 2} \over ( \bar \kappa _+ + \bar B \bar \kappa _-)^{1 \over 2}}
+ { 2 \rho ^2 \over c_6}=0
\eea
Its dependence on $\kappa _\pm$ cannot be reduced to a dependence solely on $\lambda$ since the one-form $\bar \kappa _-$ is needed to combine with $\rho^2$ to produce an  equation in which all terms transform as $(1,0)$ forms, and we obtain, 
\bea
\label{5e1a}
f^3 \, (\p_w B) \, { ( \bar \lambda + \bar B )^{3 \over 2} \over (B \bar \lambda + 1)^{1 \over 2}}
- f^3 \, (\p_w \bar B) \, { ( B \bar \lambda +  1)^{3 \over 2} \over ( \bar \lambda + \bar B)^{1 \over 2}}
+ { 2 \rho ^2 \over c_6 \, \bar \kappa _- }=0
\eea
 In summary, we have reduced the remaining BPS equations and expressed them in terms of complex differential equations given by   (\ref{5c5}) and (\ref{5e1}) along with their complex conjugate equations.  We will show in appendix~\ref{app:Bianchi-G} that these equations imply that the Bianchi identities are satisfied, and that there are hence no more constraints to take into account.
Integrating these equations will give the complex scalar field $B$ and the metric field $\rho$ in terms of the holomorphic one forms $\kappa_\pm$.   In the next section we will perform several variable redefinitions which bring the equations into a form where they can be decoupled and integrated.

\subsection{Decoupling by changing  variables {\rm (step 4/5)}}

In this subsection, we will perform two consecutive changes of variables to decouple the remaining equations. The corresponding choices will be motivated first and then carried out on the equations. 

\subsubsection{First change of variables, from \texorpdfstring{$B$}{B} to \texorpdfstring{$Z$}{Z}}

A first change of variables replaces $B$ by a complex field $Z$ and is designed to parametrize  the phase 
$e^{i \tet}$ in (\ref{eqn:Beq-lambda}) without the square root required  from its definition in (\ref{thphase}).  The following rational change of variables eliminates $B$ in terms of a complex function  $Z$ by, 
\bea
\label{8a1}
Z^2=\frac{\lambda+B}{B\bar\lambda+1}
\hskip 1in
 B=\frac{Z^2 - \lambda}{1- \bar\lambda Z^2}
\eea
and will allow us to express $e^{i \tet}$ and $f^2$ as rational functions of $Z$ and its complex conjugate,
\bea
\label{8a3}
 e^{i\vartheta}=\frac{Z}{\bar Z} \left ( \frac{1-\lambda \bar Z^2}{1-\bar\lambda Z^2} \right )
 \hskip 1in 
f^2 = { (1 - \lambda \bar Z^2) (1-\bar \lambda Z^2) \over ( 1 - |\lambda |^2) ( 1 - |Z|^4)}
\eea
The derivatives of $B$ and $\bar B$  take the following form,
\bea
\label{8a5}
\p_w B & = & { 1 - |\lambda|^2 \over ( 1 - \bar \lambda Z^2)^2} \, \p_w Z^2 - { \p_w \lambda \over 1 - \bar \lambda Z^2}
\no \\
\p_w \bar B & = & { 1 - |\lambda|^2 \over ( 1 - \lambda \bar Z^2)^2} \, \p_w \bar Z^2 + { \bar Z^2 ( \bar Z^2 - \bar \lambda) \p_w \lambda \over (1 - \lambda \bar Z^2)^2}
\eea
Implementing this change of variables on equation (\ref{eqn:Beq-lambda}) produces the following form,
\bea
\label{8b1}
&&  (2+4 |Z|^2)\partial_w Z  -Z^2 (4+2|Z|^2)\partial_w \bar Z
\no \\ && \hskip 0.3in
=  \frac{2 \bar Z (1+|Z|^2+|Z|^4)-\bar\lambda Z (1+4|Z|^2+|Z|^4)}{1-|\lambda|^2} \, \p_w \lambda
\eea
Equation (\ref{eqn:Beq-lambda}) was just one linear combination of the two equations in  (\ref{5c5}), given by the difference of the two equations (\ref{5c5}). We will take the first equation of  (\ref{5c5}) as the complimentary independent equation, and eliminate its $B$-dependence in favor of $Z$,
\bea
\label{8b2}
\p_w \ln \left ( { \rho ^2 \over \kappa _- \bar \kappa _-} \, { | 1 - Z^2 \bar \lambda| (1-|Z|^2) \over f |Z| (1-|\lambda|^2)(1+|Z|^2) } \right )
= \half \p_w \ln { Z \over \bar Z} - { \bar Z \over Z } \, \left ({ \p_w \lambda \over 1 - |\lambda|^2}\right )
\eea
Finally, eliminating $B$ in favor of $Z$ in the algebraic flux equation (\ref{5e1a}) as well, we obtain,
\bea
\label{8d6}
(1-|\lambda|^2) \p_w \left ( { Z^2 + \bar Z^{-2} \over 1- |\lambda |^2} \right )
- { 2 \p_w \lambda \over 1-|\lambda |^2} + 2  \hat \rho^2 \kappa _- \, { |Z| \over \bar Z^3} \, (1 + |Z|^2)^2=0
\eea
It remains to solve the system of equations (\ref{8b1}), (\ref{8b2}) and (\ref{8d6}).

\subsubsection{Second change of variables, from \texorpdfstring{$Z$ to $R, \psi$}{Z to R, psi}}

A second change of variables is inspired by the form of equation (\ref{8b2}), in which the norm of $Z$ and its phase enter in distinct parts of the equation. We express the complex field $Z$ in terms of two real variables, the absolute value $R$ of $Z$, and its phase $\psi$, 
\be
\label{8d1}
Z^2=R \, e^{i \psi }  
\ee
It will also be useful to change variables from  $\rho$ to $\hat \rho$ in the following way,
\be
\hat \rho^2 =
{ \rho ^2 \over c_6\kappa _- \bar \kappa _-} \, { | 1 - Z^2 \bar \lambda| (1-|Z|^2) \over f |Z| (1-|\lambda|^2)(1+|Z|^2) }
\ee
In terms of these variables (\ref{8b2}) takes the form,
\bea
\label{8d2}
\p_w \ln \hat \rho ^2 - { i \over 2} \p_w \psi + e^{- i \psi} \, { \p_w \lambda \over 1 - | \lambda |^2} =0
\eea
while  (\ref{8b1}) becomes,
\bea
\label{8d3}
(1-R^2) { \p_w  R \over R}  +  ( 1 + 4 R +R^2) \left ( i \p_w \psi + { \bar \lambda \p_w \lambda \over 1 - |\lambda|^2} \right )
- { 2 e^{-i \psi} (1+R+R^2)  \over 1 - |\lambda |^2} \, \p_w \lambda =0
\qquad
\eea
and (\ref{8d6}) becomes, 
\bea
\label{8d7}
\left ( R -{ 1 \over R} \right ) \p_w R + (R^2+1) \left ( i \p_w \psi +  { \bar \lambda \p_w \lambda \over 1 -|\lambda |^2} \right)
- { 2 R \p_w \lambda \over 1 - |\lambda |^2} \, e^{-i \psi}
+ 2 \hat \rho^2 \kappa _- e^{i \psi /2} (1+R)^2=0
\no \\
\eea
The three equations (\ref{8d2}), (\ref{8d3}), and (\ref{8d7}) are the basic starting point for the complete solution of the full system of reduced BPS equations.

\subsubsection{Decoupling the equations for \texorpdfstring{$\psi$ and $\hat \rho^2$}{psi and rho}}

Adding (\ref{8d3}) and (\ref{8d7}) cancels the terms proportional to $\p_w R$. Remarkably, and the secret to decoupling the equations, is now that the entire $R$-dependence of this sum resides in an overall multiplicative factor of $(1+R)^2$.
Omitting this factor, the sum becomes,
\bea
\label{8e1}
2i \p_w \psi +  { 2 \bar \lambda \p_w \lambda \over 1 -|\lambda |^2} - { 2  \p_w \lambda \over 1 - |\lambda |^2} \, e^{-i \psi}
+ 2 \hat \rho^2 \kappa _- e^{i \psi /2}=0
\eea
Together with (\ref{8d2}), which we repeat here for convenience,
\bea
\label{8e2}
\p_w \ln \hat \rho ^2 - { i \over 2} \p_w \psi +  { \p_w \lambda \over 1 - | \lambda |^2} \, e^{- i \psi} =0
\eea
equation (\ref{8e1}) forms a system of equations for only two of the three real fields of the system, namely $\psi $ and $\hat \rho^2$. This system is very similar to the one solved in \cite{D'Hoker:2007xy}, and we will approach it with similar methods. We note in passing that the integrability condition on (\ref{8e2}) viewed as an equation for $\hat \rho^2$ as a function of $\psi$ and given $\lambda$ is given by, 
\bea
2 \p_w \p_{\bar w} \psi + \p_{\bar w} \left (   { 2 i \p_w \lambda \over 1 - | \lambda |^2} \, e^{- i \psi} \right )
-
\p_w \left (  { 2 i \p_{\bar w} \bar  \lambda \over 1 - | \lambda |^2} \, e^{+ i \psi} \right ) =0
\eea
which is a conformal invariant field equation for $\psi$ of the sine-Gordon Liouville type \cite{D'Hoker:1982er}, very similar to equation (1.3) of \cite{D'Hoker:2007xy}. The equation for $R$ will be dealt with in section \ref{secRsol}.

\sm

Adding twice (\ref{8e2}) to (\ref{8e1}) eliminates the term proportional to $e^{ - i \psi}$, and we obtain,
\bea
\label{8e3}
 \p_w \ln \hat \rho ^2  + { i \over 2}  \p_w \psi - \p_w \ln ( 1 -|\lambda |^2)
+  \hat \rho^2 \kappa _- e^{i \psi /2}  =0
\eea
Clearly, this equation involves only the following specific complex combination of $\hat \rho^2$ and $\psi$,
\bea
\label{8e4}
\xi = { 1 \over \hat \rho^2 \, e^{ i \psi /2} }
\eea
in terms of which we may express (\ref{8e3})  as follows,
\bea
\label{8e6}
\p_w \Big ( \xi ( 1 - | \lambda |^2 ) \Big ) = \kappa _- (1 - |\lambda |^2 ) = \kappa _- - \kappa _+ \bar \lambda
\eea
where we have used the relation $\kappa _+ = \lambda \kappa _-$.

\sm

With the help of this sequence of changes of variables, the integrable structure of the system of  equations (\ref{8b1}), (\ref{8b2}) and (\ref{8d6}) has been brought out clearly. Indeed, equation (\ref{8e6}) involves only the field $\xi$, which is the particular combination of $\hat \rho$ and $\psi$ used in (\ref{8e4}). Having solved for $\xi$, either equation (\ref{8e2}) or (\ref{8e3}) becomes an equation for only a single variable, either $\hat \rho$ or $\psi$, and may be solved. Finally, having $\hat \rho$ and $\psi$, equation (\ref{8d3}) becomes an equation for $R$ only, and we will see below that it may be solved as well.

\subsection{Solving  for \texorpdfstring{$\psi$, $\hat \rho^2$ and $R$ in terms of  $\cA_\pm$ {\rm (step 5/5)}}{psi and rho}}

Having decoupled the reduced BPS equations in the preceding subsection, we will solve the decoupled equations in the present section. To do so, we begin by solving equation (\ref{8e6}) for $\xi$, and then obtain $\psi, \hat \rho^2$ and $R$.  Introducing locally holomorphic functions $\cA_\pm$ such that,
\bea
\label{8e7}
\p_w \cA_\pm = \kappa _\pm
\eea
the function $\lambda$ may be expressed in terms of $\cA_\pm$ by,
\bea  
\lambda = { \p_w \cA_+ \over \p_w \cA_-}
\eea 
Given the one-forms $\kappa_\pm$, the functions $\cA_\pm$ are unique up to an additive constant for each function. Viewed as equations for the supergravity fields in terms of $\cA_\pm$, the reduced BPS equations are therefore invariant under shifting the holomorphic functions $\cA_\pm $ by arbitrary complex constants.

\subsubsection{Solving for \texorpdfstring{$\xi$}{xi}}

In terms of $\cA_\pm$ equation (\ref{8e6}) may be recast in the following form,
\bea
\label{8e8}
\p_w \Big ( \xi (1-\lambda \bar \lambda) - \cA_- + \cA_+ \bar \lambda \Big ) =0
\eea
Equation (\ref{8e8}) is solved in terms of an arbitrary locally holomorphic function $\cA_0$ by,
\bea
\label{8e9}
\xi ( 1 - \lambda \bar \lambda) = \cA_- - \cA_+ \bar \lambda + \bar \cA_0
\eea
which provides the general solution to (\ref{8e6}). To determine $\cA_0$ in terms of $\cA_\pm$, we
enforce  (\ref{8e2}) on the result (\ref{8e9}). Upon eliminating $\xi$ using (\ref{8e9}), we find,
\bea
(1 - \lambda \bar \lambda)  \p_w \cA_0   =
(\p_w \lambda) \Big ( \bar \cA_+ + \bar \cA_0 + \cA_- -  \bar \lambda ( \cA_+ + \cA_0 + \bar \cA_-  ) \Big )
\eea
To proceed we change variables from the holomorphic function $\cA_0$ to a new holomorphic function $\f$, related as follows, $\cA_0 = - \cA_+ + \lambda (\cA_- + \f)$. The equation for $\f$ then becomes,
\bea
\label{8e11}
(1-\lambda \bar \lambda) \p_w \f +\f \, \p_w \ln \lambda = \bar \lambda \bar \f \, \p_w \ln \lambda
\eea
If $\lambda$ is constant, then $\f$ must be constant as well. Assuming henceforth that $\p_w \lambda \not= 0$, we take the derivative of the entire equation with respect to $\bar w$, and regroup terms according to their holomorphicity properties,
\bea
- { \lambda \p_w \f \over \p_w \ln \lambda} = { \p_{\bar w} (\bar \lambda \bar \f) \over \p_{\bar w} \bar \lambda}
\eea
The left side is holomorphic, while the right side is anti-holomorphic, and hence both sides must equal a complex constant $\alpha$, so that we find,
\bea
\label{8e12}
\lambda \p_w \f & = & - \alpha \, \p_w \ln \lambda
\no \\
\p_w ( \lambda \f) & = & \bar \alpha \, \p_w \lambda
\eea
Eliminating the derivative $\p_w \f$ between both equations gives,
\bea
\f = { \alpha \over \lambda } + \bar \alpha
\eea
Assembling all these results gives,
\bea
\xi (1-\lambda \bar \lambda) = (\cA_- - \bar \cA_+ + \bar \alpha) + \bar \lambda(   \bar \cA_- - \cA_+ + \alpha)
\eea
Recalling that the functions $\cA_\pm$ were defined only up to additive constant shifts, we may absorb the constant $\alpha$ into $\cA_\pm$, so that our final expression for the solution becomes, 
\be
\label{calA}
\xi = {\cL \over  1- |\lambda |^2}
\ee
where $\cL$ is given by,
\be
\label{calA2}
\cL =  (\cA_- - \bar \cA_+ ) + \bar \lambda(   \bar \cA_- - \cA_+)
\eea
Note that   $\hat \rho$ and $\psi$  are directly determined by $\xi$ using  equation (\ref{8e4}).

\subsubsection{Solving  for \texorpdfstring{$R$}{R}} 
\label{secRsol}

To solve for $R$, we start from equation (\ref{8d3}) and eliminate the term proportional to $e^{-i \psi}$ using (\ref{8d2}). We then divide the resulting equation by $R$, and find,
\bea
\label{Reqa}
0 & = &
\left ( { 1 \over R^2} -1 \right ) \p_w R + \left ( R+{1 \over R} +4 \right ) \left ( i \p_w \psi 
- \p_w \ln (1- | \lambda |^2 ) \right )
\no \\ && \quad
+ \left ( R + { 1 \over R} +1 \right ) ( 2 \p_w \ln \hat \rho^2 - i \p_w \psi)
\eea
Changing variables from $R$ to $W$ defined by,
\bea
W=R + { 1 \over R}
\eea
renders   equation (\ref{Reqa}) linear in $W$, with an inhomogeneous part,
\bea
\label{Xeq}
\p_w W -2 (W+1) \p_w \ln \hat \rho^2 + (W+1) \p_w \ln (1 - \lambda \bar \lambda) = 3 i \p_w \psi - 3 \p_w \ln (1-\lambda \bar \lambda)
\eea
The homogeneous equation is solved straightforwardly by,
\bea
W + 1= W_0 \, { \hat \rho^4 \over 1 - \lambda \bar \lambda}
\eea
To solve the homogeneous equation, $W_0$ is an arbitrary constant which is required to be real for real $W$.
To find a particular solution to the inhomogeneous equation, we let $W_0$ be  a function, which then must satisfy,
\bea
\p_w W_0 = { 1 - \lambda \bar \lambda \over \hat \rho^4} \, 3 i \p_w \psi - 3 { \p_w (1-\lambda \bar \lambda) \over \hat \rho^4}
\eea
Recasting the equation in terms of the variable $\xi = \hat \rho^{-2} e^{i \psi /2}$, and then expressing $\xi$ in terms of $\cL$ and $\lambda$, using (\ref{calA}), we find, 
\bea
\p_w W_0 =  { 3 \cL \bar \cL \over 1 -\lambda \bar \lambda} \, \p_w \ln { \bar \cL \over \cL}
-{ 3 \p_w (\cL \bar \cL) \over 1 -\lambda \bar \lambda}
+   \p_w \left ( { 3 \cL \bar \cL  \over 1- \lambda \bar \lambda} \right )
\eea
The first two terms on the right side may be evaluated using the expression for $\cL$ of (\ref{calA2}). Putting all together, we find the following equation for $W_0$, 
\bea
\p_w \left ( W_0 - { 3 \cL \bar \cL  \over 1- \lambda \bar \lambda} \right )
= - 6 \bar \cL \p_w \cA_-
\eea
Using the explicit formula for $\bar \cL$, and the fact that $\lambda \p_w \cA_- = \p_w \cA_+$, we find,
\bea
\p_w \left ( W_0 - { 3 \cL \bar \cL  \over 1- \lambda \bar \lambda} - 6 \cA_+ \bar \cA_+ + 6 \cA_- \bar \cA_- \right )
=  6 (\cA_+ \p_w \cA_- - \cA_- \p_w \cA_+ )
\eea
Now the right side is a holomorphic 1-form, and so locally there exists a holomorphic function $\cB$, defined up to the addition of an arbitrary complex constant such that,\footnote{Clearly, the holomorphic function $\cB$ should not be confused with the 10-dimensional charge conjugation matrix of footnote 3 for which the same symbol is being used here.}
\bea
\label{calB}
\cA_+ \p_w \cA_- - \cA_- \p_w \cA_+ = \p_w \cB
\eea
The general solution is then given by,
\bea 
\label{X0}
W_0 & = &  { 3 \cL \bar \cL  \over 1- \lambda \bar \lambda} + 6 \cA_+ \bar \cA_+ -  6 \cA_- \bar \cA_- +  6\cB +  6 \bar \cB
\eea
and an arbitrary integration constant, which  parametrizes the admixture of the solution to the homogeneous equation, has been absorbed into $\cB$. Note that the entire solution is real, as is required by the nature of $W$ and $R$. This completes the solution of the decoupled reduced BPS equations for the fields $\psi, \hat \rho$ and $R$.

%%%%%%%%%%%%%%%%%%%%%%%%%%%%%%%%%%%%%%%%%%%
%%%%%%%%%%%%%%%%%%%%%%%%%%%%%%%%%%%%%%%%%%%
\section{Local solution to Type IIB supergravity}
\label{sec:5}
\setcounter{equation}{0}
%%%%%%%%%%%%%%%%%%%%%%%%%%%%%%%%%%%%%%%%%%%
%%%%%%%%%%%%%%%%%%%%%%%%%%%%%%%%%%%%%%%%%%%

In this section we summarize the complete local solution for the supergravity fields, which is parametrized by two holomorphic functions $\cA_\pm$ and various constants. The doublet $(\cA_+, \cA_-)$ transform linearly  under the group $SU(1,1)$, which is isomorphic to the group SL(2,$\mathds{R}$).  The transformation properties of the supergravity fields under $SU(1,1)$  can be made transparent by expressing the supergravity fields with the help of natural invariants. The $SU(1,1)$ transformation properties of the holomorphic data and its invariants  will be spelled out in  section~\ref{sec:5-1}, before we give the supergravity fields in \ref{sec:5-2} and discuss the $SU(1,1)$ transformations induced on them in section \ref{sec:5-3}. We discuss positivity and regularity conditions respectively in sections \ref{sec:5-4} and \ref{sec:5-5}. In \ref{sec:5-6} we show how the T-dual of the D4/D8 solution in massive type IIA supergravity can be recovered as a special case of our general solution, while in \ref{sec:5-7} we give a local solution which satisfies the positivity and regularity conditions in a finite but local region near a boundary segment. Finally,  in section \ref{sec:5-8}, we discuss the conditions on the holomorphic data $\cA_\pm$ and on the supergravity fields under which solutions with monodromy can exist.

\subsection{\texorpdfstring{$SU(1,1)$}{SU(1,1)} transformations of the holomorphic data }
\label{sec:5-1}

The basic data parametrizing the general local solution are two holomorphic functions $\mathcal A_\pm$,
which transform linearly under $SU(1,1)$. Parametrizing the elements of $SU(1,1)$ by  $u, v\in\mathds{C}$ subject to $|u|^2-|v|^2=1$, the functions $\cA_\pm$ transform as,
\bea
\label{eq:5-1}
 \cA_+ & \rightarrow & \cA_+^\prime = u \cA_+ - v \cA_-
 \no \\
 \cA_- &\rightarrow & \cA_-^\prime = \bar u \cA_- - \bar v \cA_+
\eea
Note that the $\bar\cA_\mp$ and $\kappa _\pm$ transform in the same fashion as $\cA_\pm$ does.
The functions $\cA_\pm$ determine the holomorphic function $\cB$ introduced in (\ref{calB}), up to an additive constant, by,
\begin{align}
 \partial_w \cB&= \cA_+ \p_w \cA_- - \cA_- \p_w \cA_+
\end{align}
The right hand side is invariant under the $SU(1,1)$ transformations (\ref{eq:5-1}) and consequently $\mathcal B$  transforms at most by a constant shift. Derived quantities which will copiously appear in the expressions for the supergravity fields are $\kappa_\pm$ and $\lambda$ defined by,
\begin{align}
 \kappa_\pm&=\partial_w \mathcal A_\pm
 &
 \kappa_+&=\lambda\kappa_-
\end{align}
Since  $\kappa_\pm$ transform as $\cA_\pm$, we are led to the following natural invariant,
\bea
\kappa^2  =-|\kappa_+|^2+|\kappa_-|^2
\eea
Finally, a combination which was already encountered earlier in (\ref{X0}) and given by,
\bea
\cG = |\cA_+ |^2 - | \cA_- |^2 + \cB + \bar\cB
\eea
also has simple transformation properties, since the first and second terms on the right side combine into an $SU(1,1)$ invariant. Since $\cB$ transforms at most by shifts  under $SU(1,1)$, so does $\mathcal G$.
Moreover, $\cG$ is the only place where $\cB$ shows up, and we see that of the generally complex constant of integration hiding $\cB$, only the real part is relevant. For later use, we note the following relation between $\cG$ and $\kappa ^2$ given by $\kappa^2=-\partial_w\partial_{\bar w}\cG$.

\subsection{Supergravity fields in terms of holomorphic data}
\label{sec:5-2}

The general local solution to Type IIB supergravity with $SO(2,5)\times SO(3)$ symmetry can now be expressed in terms of the holomorphic data introduced above. Translating the local solution to the reduced BPS equations obtained in section~\ref{sec:4} back to the supergravity fields is straightforward, except in the case of the flux field for which the derivation of the flux potential is quite involved, and detailed calculations are relegated to appendix~\ref{sec:10}. Here we will summarize the results and discuss some of the immediate properties of the solutions.

\sm

Recall that the symmetries of the problem dictate the Ansatz for the bosonic supergravity fields, while the fermionic fields vanish. The five-form field strength vanishes, while the Ansatz for the metric and three-form flux fields are as follows, 
\bea
ds^2 & = &  f_6^2 \, ds^2 _{\mathrm{AdS}_6} + f_2 ^2 \, ds^2 _{\mathrm{S}^2} + ds^2 _\Sigma
\hskip 1in 
ds^2_\Sigma=4\rho^2 dw d\bar w
\no \\
F_{(3)} & = & d C_{(2)}  \hskip 2.4in C_{(2)} = \cC  \, \hat e^{67}
\eea
where $f_6, f_2, \rho, \cC$ and the dilaton-axion field $B$ are all functions on $\Sigma$, and $\hat e^{67}$ is the volume form on an S$^2$ of unit radius defined in (\ref{frame2}).

\sm

The metric functions $f_2$, $f_6$ and $\rho$ can be expressed solely in terms of $\cG$,
and for notational convenience we introduce the composite quantities $R$ and $W$ defined in terms of $\cG$ through,
\begin{align}
\label{eq:5-2a}
 W=R+\frac{1}{R}&=2+6\,\frac{\kappa^2 \, \cG }{|\partial_w\cG|^2}
\end{align}
The right  side is real and so are $R$ and $W$.  The metric functions are then given by,
\begin{align}
\label{eq:5-2b}
 f_2^2&=\frac{c_6^2  \kappa^2 (1-R)}{9 \, \rho^2 (1+R)}
 &
 f_6^2&=\frac{c_6^2 \kappa^2 (1+R) }{\rho^2 \, (1-R)}
 &
 \rho^2&=\frac{  c_6(R+R^2)^\half }{|\partial_w \cG|} \left (  \frac{\kappa ^2 }{1-R} \right )^{3 \over 2} 
\end{align}
The axion-dilaton field $B$ of (\ref{Btau})  is given by,
\begin{align}
\label{eq:5-2-Bfield}
 B&=\frac{\kappa_+\partial_{\bar w} \cG - \bar\kappa_- R\, \partial_w \cG}
 {\bar\kappa_+ R\, \partial_w\cG -  \kappa_- \partial_{\bar w}  \cG}
\end{align}
Finally, the flux potential function $\cC$ for the three-form field strength $F_{(3)}$  is derived in appendix \ref{sec:10} and given by, 
\begin{align}
\label{eq:5-2c}
 \cC &= \frac{4 i c_6}{9}\left (  \frac{\bar\kappa_- W \partial_w\cG-2\kappa_+\partial_{\bar w}\bar \cG}{(W+2) \, \kappa^2 }  - \bar  \cA_- - 2 \cA_+ - \cK_0 \right )
\end{align}
Here, $\cK_0$ is a complex integration constant which represents the residual gauge transformation degree of freedom in $C_{(2)}$ restricted to our Ansatz, and does not affect the gauge-invariant field strength $F_{(3)}$.

\subsection{\texorpdfstring{$SU(1,1)$}{SU(1,1)} transformations induced on the supergravity fields}
\label{sec:5-3}

The expressions for the supergravity fields in terms of the holomorphic data, obtained in section \ref{sec:5-2} allow us to specify more precisely which transformations of the holomorphic data leave the supergravity fields invariant, or transform them according to $SU(1,1)$-duality. The radii $f_6$ and $f_2$ must be invariant since they parametrize  the metric in Einstein frame. Therefore, $R$ and the combination $\kappa^2/\rho^2$ must be invariant, and hence $|\p_w \cG|^2/\kappa ^2$ and as a result also $\cG$ and $\kappa ^2$ themselves must be invariant by (\ref{eq:5-2a}). We will now implement these invariance requirements on the holomorphic data $\cA_\pm$ themselves. 

\sm

Invariance of $\kappa ^2$ requires that the holomorphic one-forms transform under $SU(1,1)$, 
\bea
\kappa_+ & \to &  \kappa _+ '= + u \kappa_+ - v \kappa_-
\no \\
\kappa_- & \to & \kappa _- '= - \bar v \kappa_+ + \bar u \kappa_-
\eea
where we have parametrized $SU(1,1)$ by $u,v\in \CC$ with $|u|^2-|v|^2=1$. Integrating the above transformation laws to obtain the holomorphic functions $\cA_\pm$ we get, 
\bea
\label{Atrans}
\cA_+ & \to & \cA_+'=  + u \cA_+ - v \cA_- + a_+ 
\no \\
\cA_- & \to & \cA_-' =   - \bar v \cA_+ + \bar u \cA_- + a_-
\eea
where $a_\pm$ are complex constants. The addition of the constants $a_\pm$ leaves $\kappa _\pm$ unchanged but  transforms $\p_w\cB$ as follows,
\bea
\p_w \cB & \to & \p_w \cB' = \p_w \cB + a_+ \p_w \cA_-' - a_- \p_w \cA_+'
\eea
Integrating this relation, we find, 
\bea
\cB & \to & \cB ' = \cB + a_+ \cA_-' - a_- \cA_+' + b_0
\eea
where $b_0$ is a complex constant. Therefore, the transformation law for $\cG$ is as follows,
\bea
\cG & \to & \cG' = \cG + (a_+ - \bar a_-) (\bar \cA_+' + \cA_-') + (\bar a_+ -a_-) ( \cA_+' + \bar \cA_-')
\no \\ && \hskip 0.5in
+ a_+ \bar a_+  - a_- \bar a_-  + b_0 + \bar b_0
\eea
Invariance of $\cG$  requires,
\bea
a_- = \bar a_+ \hskip 1in b_0 + \bar b_0 =0
\eea
Since $\cA_- $ and $\bar \cA_+$ transform under $SU(1,1)$ by the same formula, the restriction $a_- = \bar a_+$ is automatically $SU(1,1)$-invariant. To analyze the transformation properties of the dilaton-axion field $B$, we use the solution (\ref{eq:5-2-Bfield}) and again appeal to the fact that  $\cG$, $\partial_w\cG$ and $R$ are invariant under $SU(1,1)$. The transformation property of $B$ under (\ref{Atrans}) is as follows,
\bea
 B \rightarrow B'=\frac{uB+v}{\bar v B+\bar u}
\eea
Finally, the transformation law for the flux potential $C_{(2)} = \cC \, \hat e^{67}$ under (\ref{Atrans}) is given by, 
\bea
 \cC \rightarrow \cC ' = u \cC +v\bar \cC -\frac{4i c_6}{9}\big(\cK_0^\prime-u\cK_0-v\bar\cK_0 + 3 a_+ \big)
\eea
The last term is constant and amounts to a gauge transformation on the field $C_{(2)}$. Therefore, up to gauge transformations, we recover (\ref{stransf}), as required. We see that the transformations of 
(\ref{eq:5-1})  indeed induce the appropriate $SU(1,1)$ transformations on the supergravity fields. But at the level of the holomorphic data we have an  additional shift symmetry, which leaves the supergravity fields 
invariant and results from the fact that $\cA_\pm$ are determined from $\kappa_\pm$ only up to additive constants.

\subsection{Positivity conditions}
\label{sec:5-4}

By construction in (\ref{eq:5-2b}), the metric fields $f_2^2, f_6^2$ and $\rho^2$ are real, but they are not necessarily positive, as is required by the Minkowski signature of ten-dimensional space-time. In this subsection, we investigate the requirements on the holomorphic data implied by the positivity of $f_2^2, f_6^2$ and $\rho^2$, and the condition $|B|\leq 1$. There are no reality or positivity requirements derived from the flux field $C_{(2)}$ since it is complex. The further conditions needed to produce regular solutions will be investigated in the next section. 

\sm

Positivity of the expressions for $f_2^2$ and $f_6^2$ in (\ref{eq:5-2b}) requires,
\bea
\label{ineq1}
\kappa^2(1-R) \geq 0
\eea
The expression for $\rho^2$ in (\ref{eq:5-2b})  is then automatically positive. By definition $R$ is an absolute value, so that we must have $R  \geq 0$ and therefore $W\geq 2$ by (\ref{eq:5-2a}), which implies,
\bea
\label{ineq2}
\kappa ^2 \, \cG \geq 0
\eea
Finally, we verify that $|B|\leq 1$ holds, or equivalently $f^2 \geq 1$,   using (\ref{8a3}),
\bea
\label{fB}
f^2 = { 1 \over 1 - |B|^2} = 1 + { |\lambda - Z^2|^2 \over ( 1-|\lambda|^2) (1-|Z|^4)}
\eea
It is manifest that $f^2 \geq 1$  since $\kappa ^2 (1-R)=|\kappa_-|^2 (1-|\lambda|^2)(1-|Z|^2)$, which is positive in view of (\ref{ineq1}). So the conditions (\ref{ineq1}) and (\ref{ineq2}) exhaust the reality constraints.  As a result, there are two branches to the solutions, 
\bea
&& \left \{ \kappa ^2 \geq 0, \qquad R \leq 1, \qquad \cG \geq 0 \right \}
\no \\
&& \left \{ \kappa ^2 \leq 0, \qquad R \geq 1, \qquad \cG \leq 0   \right \}
\eea
Actually, these two branches are mapped into one another under a complex conjugation, which includes the reversal of the complex structure on $\Sigma$. Specifically, this transformation reverses the sign of $\kappa ^2$ and $\cG$, maps $R $ to $R^{-1}$, and interchanges $w$ and $\bar w$, and may be realized on the holomorphic functions as follows,
\bea
\cA _+ (w) & \to & \cA '_+ (w) = \bar \cA _- (w) = \overline{ \cA _- (\bar w)}
\no \\
\cA _- (w) & \to & \cA '_- (w) = \bar \cA _+ (w) = \overline{ \cA _+ (\bar w)}
\eea
and thus on holomorphic forms by,
\bea
\kappa _+ (w) & \to & \kappa '_+ (w) = \bar \kappa _- (w) = \overline{ \kappa _- (\bar w)}
\no \\
\kappa _- (w) & \to & \kappa '_- (w) = \bar \kappa _+ (w) = \overline{ \kappa _+ (\bar w)}
\eea
One verifies that these transformations have the desired action on $\kappa ^2$, $\cG$, and $R $, and therefore leave the metric functions $f_2^2, f_6^2, \rho^2$ invariant, while complex conjugating the fields $B$ and $\cC$ combined with a reversal of the complex structure on $\Sigma$,
\bea
B(w,\bar w) & \to & B'(w,\bar w) = \bar B (w,\bar w) = \overline{B(\bar w, w)}
\no \\
\cC (w,\bar w) & \to & \cC '(w,\bar w) \, = \bar \cC (w,\bar w) \, = \overline{\cC(\bar w, w)}
\eea
Therefore, we may restrict to considering just a single branch of the solutions, specified by,
\bea
\label{poscon}
\kappa ^2 \geq 0 \hskip 1in  R \leq 1 \hskip 1in \cG \geq 0
\eea
the other branch being related by complex conjugation.

\subsection{Regularity conditions}
\label{sec:5-5}

To describe holographic duals to $4\,{+}\,1$ dimensional CFTs, we are mainly interested in solutions where the $AdS_6$ factor governs the entire non-compact part of the geometry. Therefore, we will assume that $\Sigma$ is compact, with or without boundary. It will be convenient to examine the regularity conditions required in each one of these two cases separately.

\sm

When the Riemann surface $\Sigma$ is compact and without boundary, a regular supergravity solution requires the metric functions $f_2^2, f_6^2$ and $\rho^2$ to remain strictly positive, and the axion-dilaton field to satisfy the strict inequality $|B| <1$  throughout $\Sigma$. As a result, the  corresponding conditions on the holomorphic data are given by the strict inequalities,
\bea
\kappa ^2 > 0 \hskip 1in  R < 1 \hskip 1in \cG > 0
\eea
throughout $\Sigma$. The  inequality $\kappa ^2 >0$ implies the strict inequality  $|\lambda | <1$. 

\sm

When the Riemann surface $\Sigma$ has a non-empty boundary $\p \Sigma$, a regular supergravity solution may be obtained when the function $f_2^2$ vanishes on $\p \Sigma$, provided such behavior corresponds to the shrinking of a sphere $S^2$ as part of a regular  three-dimensional sub-manifold. It is clear from the explicit solutions in (\ref{eq:5-2b})  that $9 f_2^2 < f_6^2$ and therefore the vanishing of $f_6^2$ will force $f_2^2$ to vanish as well producing a space-time geometry with a short-distance singularity. To avoid such physically unacceptable singularities, we will assume henceforth that the regular part of the  boundary of $\Sigma$ is characterized by $f_2^2=0$ and $f_6^2>0$. In particular, a topologically non-trivial three-cycle, or three-sphere, may arise as part of the space-time manifold of a regular solution from fibering the sphere $S^2$ over a line segment on $\Sigma$ which is spanned between two points on $\p \Sigma$, and which cannot be continuously contracted to a point. 

\sm

We will now investigate the behavior of a regular supergravity solution, and its associated holomorphic data, near a boundary point or segment of $\Sigma$.   To analyze this behavior, we express $f_2$ and $f_6$ as directly as possible in terms of holomorphic data, and we find,
\bea
{f_2^2 \over f_6^2} = \frac{(1-R)^2}{9 (1+R)^2}
\hskip 1in
{ f_6^2 \over  c_6} =   \left | \p_w \cG \right | 
\left ( { 1 - R \over \kappa ^2} \right )^\half \left  (1+{1 \over R} \right )^\half
\eea
The neighborhood of a regular boundary point is realized by letting $f_2 ^2 \to 0$ while keeping $f_6^2$ finite. 
On the variables $R$ and $\kappa^2$, this requires the following limiting behavior,  
\bea
1-R \to 0 \hskip 1in { 1-R \over \kappa ^2}\to \hbox{finite}
\eea
The first condition ensures that $f_2/f_6\rightarrow 0$, the second that $f_6$ stays finite. As a result, 
we have $\kappa ^2 \to 0$ and $R \to 0$  at a regular boundary point, while their ratio stays finite.

\sm

Furthermore, the limiting behavior $R\to1$ implies that $W\to 2 + \cO((R-1)^2)$, which imposes a condition on the behavior of $\cG$ in view of the equation (\ref{eq:5-2a}) for $R$. To derive this condition,  we recast  (\ref{eq:5-2a}) as follows,
\bea
{ 1-R \over  \kappa ^2} = { 6 R \, \cG \over (1-R) |\p_w \cG|^2}
\eea
Near a regular boundary point, the left side remains finite. Assuming that $\p_w \cG$ also remains finite, we are led to the following limiting behavior for $\cG$, 
\bea
{ \cG \over 1-R} \to \hbox{finite}
\eea
Finally, we examine the regularity condition on the axion-dilaton field, namely $|B|<1$ or equivalently $f^2 \geq 1$ and $f^2$ remains finite near a regular boundary point of $\p \Sigma$. From equation (\ref{fB}) these conditions will be realized provided the following ratio has a finite limit, 
\bea
{Z^2 -  \lambda  \over 1-R} & \to & \hbox{finite}
\eea
which gives us information on the behavior of the relative phase between $\lambda$ and $Z^2$.
To analyze this condition, we make use of the definition $Z^2=R e^{i \psi}$ and the relation $e^{i \psi} = \bar \xi/\xi= \bar \cL / \cL$, with $\cL$ defined in (\ref{calA2}). Expressing the result further with the help of $\cA_\pm$ we obtain,  
\bea
{Z^2 -  \lambda  \over 1-R} = {\bar \cA_- - \cA_+ \over \cL} \times { 1-|\lambda|^2  \over 1-R}
& \to & \hbox{finite}
\eea
This condition is automatically fulfilled as long as $(\bar \cA_- - \cA_+)/  \cL$ remains finite.

\sm

In summary, the behavior of the supergravity fields near a  boundary point on $\p \Sigma$ is regular if and only if  $\kappa ^2 \to 0$ and the following two ratios have a finite and non-zero limit,
\bea
\label{regcon}
{ 1-R \over \kappa ^2} \to \hbox{finite}
\hskip 1in 
{ \cG \over \kappa ^2 } \to \hbox{finite}
\eea
When these conditions are obeyed, all supergravity fields are regular in the neighborhood of the corresponding regular boundary point. Of course, one may wish to consider supergravity solutions with sufficiently mild singularities, such as a diverging dilaton field at isolated points on the boundary. In this case, the condition $\cG /\kappa ^2$ being finite may have to be relaxed.

\subsection{Recovering the T-dual of D4/D8 }
\label{sec:5-6}

In this subsection, we will recover the T-dual of the D4/D8 solution given in \cite{Lozano:2013oma} and show
that this solution, even though singular, solves the BPS equations derived in the present paper.
To find the T-dual of D4/D8, we consider a local coordinate system $w, \bar w$ on a Riemann surface $\Sigma$, and make the following Ansatz for the holomorphic data,
\bea
\label{eq:5-4a}
 \cA_\pm = \frac{a}{2}w^2 \mp b w
\eea
with $a,b\in\mathds{R}$. As a result, we find the following auxiliary quantities,
\bea
\kappa_\pm=aw \mp b
\hskip 0.8in 
\kappa^2=2ab(w+\bar w)
\hskip 0.8in 
 \cB=\frac{ab}{6}(1-2w^3)
\eea
The combinations $\cG$ and $R$, which are also required to construct the supergravity fields of the solutions,  are given as follows, 
\begin{align}
 \cG&=\frac{ab}{3Y^2}
 &
 R&=\frac{Y-1}{Y+1}
 &
 Y&=\frac{1}{\sqrt{1-(w+\bar w)^3}}
\end{align}
The positivity conditions of (\ref{poscon}) require $\kappa ^2 \geq 0$, $\cG \geq 0$ and $R \leq 1$, and thus,
\bea
0 < ab 
\hskip 1in 
0 \leq w+ \bar w < 1
\eea
Note that neither $w+\bar w=0$, nor $w+\bar w=1$ satisfy the regularity conditions of (\ref{regcon}), so that we must expect the resulting supergravity solution to have singularities there.

\sm

The metric functions are found as follows,
\bea
 \rho^2=c_6\sqrt{2ab}(w+\bar w)Y^{3/2}
\hskip 0.6in
 f_6^2=c_6^2\sqrt{2ab}Y^{-1/2}
\hskip 0.6in 
 f_2^2=\frac{c_6^2}{9}\sqrt{2ab}Y^{-5/2}
\eea
For the axion and dilaton we use the formulas in appendix~\ref{sec:10}, eqs.\ (\ref{9phi}) and (\ref{9chi}),
to get them separately right away. This yields,
\begin{align}
 e^{-2\phi}&=\frac{2b Y}{a(w+\bar w)}
 &
 \chi&=\frac{ia}{2b}(w-\bar w)
\end{align}
To match to \cite{Lozano:2013oma}, we change from $w, \bar w$  to real coordinates $\theta$, $\phi_3$ defined by,
\bea
\label{real-coords}
  \cos\theta=\left(w+\overline w\right)^{3/2}
\hskip 1in
  \phi_3=\frac{ia}{2bm}(w-\overline w)
\eea
and fix the parameters $a,b, c_6$ by the following choice, 
\begin{equation}
\label{eq:5-4b}
  a=\frac{27}{16}m^{1/3}
  \qquad\quad
  b=\frac{9}{8m^{1/3}}
  \qquad\quad
  c_6=1
\end{equation}
With the notation $\widetilde W=(m\cos\theta)^{-1/6}$, the axion and dilaton fields become,
\begin{equation}
 e^{-2\phi}=\frac{3\sin\theta}{4{\widetilde W}^4}
\hskip 1in
 \chi=m\phi_3
\end{equation}
This reproduces the results of (A.1) of \cite{Lozano:2013oma}, noting that the definition
of the dilaton in (\ref{Btau}) differs from that used in \cite{Lozano:2013oma} by a factor $2$.
The metric functions become,
\begin{equation}
4 f_6^2 e^{-\phi}={\widetilde W}^2
 \qquad
4 f_2^2 e^{-\phi}={\widetilde W}^2\sin^2\theta
 \qquad
 4\rho^2 e^{-\phi} |dw|^2=
 {\widetilde W}^2\Big(d\theta^2+\frac{4d\phi_3^2}{{\widetilde W}^4\sin^2\theta}\Big)
\end{equation}
Our metric is in Einstein frame, that of \cite{Lozano:2013oma} in string frame.
The metric functions in string frame are simply given by dropping the
dilaton factors in the above expressions, i.e.\
$(f_6^2)_\mathrm{string}=e^\phi f_6^2$ etc.
Upon this change to string frame, we have exactly the metric in (A.1) of \cite{Lozano:2013oma}.
This just leaves the flux field $C_{(2)}$, which we obtain from (\ref{eq:5-2c}) as,
\begin{align}
 C_{(2)}&=\frac{i}{9}\Big(6a w\bar w-a(w+\bar w)^2(3+Y^{-2})+6b(w-\bar w)-\cK_0\Big)\hat e^{67}
\end{align}
Switching coordinates as in (\ref{real-coords}), we find,
\begin{equation}
 dC_{(2)}=\frac{5i}{8}(m\cos\theta)^{1/3}\sin^3\!\theta d\theta\wedge \hat e^{67}+(1+im\phi_3)d\phi_3\wedge \hat e^{67}
\end{equation}
To compare to \cite{Lozano:2013oma}, we separate out the real and imaginary parts, which are,
\begin{eqnarray}
 \mathrm{Re} (dC_{(2)})&=&dB_{(2)}\,=\,d\phi_3\wedge \hat e^{67}
\no \\
 \mathrm{Im}(dC_{(2)})&=& \frac{5}{8}(m\cos\theta)^{1/3}\sin^3\!\theta d\theta\wedge \hat e^{67}+m\phi_3 d\phi_3\wedge \hat e^{67}
\end{eqnarray}
The combination which appears as $F_3$ in (A.1) of \cite{Lozano:2013oma} becomes,
\begin{equation}
  \mathrm{Im}(dC_{(2)})-\chi dB_{(2)}=\frac{5}{8}(m\cos\theta)^{1/3}\sin^3\!\theta d\theta
\end{equation}
So with the choice (\ref{eq:5-4a}), (\ref{eq:5-4b}) for the holomorphic data we reproduce the T-dual of the D4/D8 solution exactly. As detailed in the introduction, this solution is singular. But the fact that we 
recover it from our general local solution provides a useful consistency  check.

\subsection{Satisfying positivity and regularity conditions locally near \texorpdfstring{$\p \Sigma$}{dSigma}}
\label{sec:5-7}

In this subsection, we shall show that, at least locally in a finite neighborhood of part of the boundary of $\Sigma$, it is possible to satisfy both positivity and regularity conditions.   By a conformal transformation, we map a boundary component of $\Sigma$ to the real axis, and take the interior of $\Sigma$ to be part of the upper half plane parametrized by complex coordinates $w, \bar w$. To realize the positivity condition $\kappa ^2=0$, or equivalently $|\lambda|=1$, on $\p \Sigma$ we choose $\lambda = (1+iw)/(1-iw)$, so that the  functions $\cA_\pm$ are related by the following equation, 
\bea
\label{12a1}
\p_w \cA_+ (w) = { 1 + i w \over 1 - i w} \, \p_w \cA_- (w) 
\eea
We restrict attention to local solutions for which $\p_w \cA_\pm$ are rational functions of $w$, and can be decomposed into a sum of simple poles plus a constant additive term, 
\bea
\p_w \cA_\pm (w) = ( 1 \pm i w) \sum _{n=1}^N { a_n \over w-x_n}
\eea
The positions of the poles $x_n$ are chosen to be real. We  guarantee the absence of zeros for $\p_w \cA_-$  in the upper half plane by requiring $ a_n \, e^{ -i \phi} \in \RR$ for all $n$ and some $n$-independent phase $\phi$. The integrals $\cA_\pm$ are as follows, 
\bea
\cA_\pm (w) = \a_\pm + \sum _{n=1}^N  a_n \Big ( (1\pm i x_n) \ln (w-x_n) \pm i w \Big )
\eea
where $\a_\pm $ are complex constants. For $N \geq 2$, the calculation of $\cB$ reveals the presence of dilogarithms. To avoid this complication, we shall examine only the simplest case $N=1$ where no dilogarithms appear in $\cB$. Setting $a_1=a$ and $x_1=x$, we then obtain, 
\bea
\kappa ^2 = { - 2 i (w-\bar w) |a|^2 \over |w-x|^2}
\eea
which indeed vanishes on $\RR$ and is strictly  positive in the upper half plane. Integrating these equations as well as the one for $\cB$, we find, 
\bea
\cA_\pm (w) & = & \a_\pm \pm i a (w-x) + a (1\pm i x) \ln (w-x)
\no \\
\cB (w) & = & \cB_0 +ia(w-x)(4a-\a_+ - \a_-) 
\no \\ &&
+  a \Big ( \a_+ (1-ix) - \a_- (1+ix) -2i a (w-x) \Big ) \ln (w-x)
\eea
where $\cB_0$ is an integration constant. We use these ingredients to compute $\cG$. The vanishing of $\cG$ on the real axis requires $\cB_0+ \bar \cB_0=0$ as well as $a^2 = - |a|^2$ and $\alpha _- = \bar \alpha _+$, and we  obtain,
\bea
\cG (w) =  - 2 i  |a|^2  ( w - \bar w) \Big (2   -  \ln | w-x|^2  \Big )  
\eea
In the upper half plane the pre-factor $-2i (w-\bar w)$ is positive, so that we have $\cG >0$ inside a semicircle centered at $x$ with $\Im (w)>0$ and  $\ln |w-x|^2 <  2$, while $\cG=0$ on the real axis. 

\sm

Therefore, we have  established that it is possible, locally in a finite region near a boundary component, to satisfy both the positivity  and regularity conditions on $\kappa ^2$ and $\cG$. On the semi-circle defined by $\ln |w-x|^2 = 2$ the behavior fails to be regular, since we have $\cG=0$ but $\kappa ^2 >0$ reminiscent of the singularity of the T-dual to D4/D8 at $w+\bar w=1$. It remains to find  supergravity solutions which obey the positivity and regularity conditions globally, an investigation that we shall reserve for future work.

\subsection{Supergravity solutions with monodromy}
\label{sec:5-8}

When $\Sigma$ is a compact Riemann surface without boundary, and the locally holomorphic functions $\cA_\pm$ are assumed to be single-valued on $\Sigma$, then they must be constant, which does not produce any supergravity solutions. Therefore, on any compact $\Sigma$ without boundary, the existence of regular supergravity solutions will require the locally holomorphic functions $\cA_\pm$ to have non-trivial monodromy, or poles which conspire in such a way that the supergravity fields remain finite.

\sm

Under the weaker assumption that the locally holomorphic function $\lambda$ is  single-valued  on a compact Riemann surface  $\Sigma$ without boundary, but $\cA_\pm$ are allowed a non-trivial monodromy, it follows from the condition $|\lambda|<1$ that $\lambda $ must be constant. The relation between $\kappa _\pm$ may then be integrated explicitly, and we have,
\bea
A_+ = \lambda A_- + a_0
\eea
where $a_0$ is an arbitrary complex constant. The monodromies of (\ref{Atrans}),
which we repeat here for convenience,
\bea
\label{Atrans1}
\cA_+ & \to & \cA_+'=  + u \cA_+ - v \cA_- + a_+ 
\no \\
\cA_- & \to & \cA_-' =   - \bar v \cA_+ + \bar u \cA_- + a_-
\eea
will be compatible with the relation $A_+ = \lambda A_- + a_0$ provided $A_+ ' = \lambda A_- ' + a_0$ as well. This condition in turn imposes two complex-valued conditions on the three complex-valued monodromy parameters $u,v,a_+=\bar a_-$, given as follows,
\bea
u \lambda - \bar u \lambda + v + \bar v \lambda ^2 & = & 0
\no \\
(1-u - \lambda \bar v) a_0 - a_+ + \lambda \bar a_+ & = & 0
\eea
Manifestly, the only solution to these equations with trivial $SU(1,1)$ monodromy, namely $u=1, v=0$ are solutions with no monodromy at all since $a_+ - \lambda \bar a_+=0$ implies $a_+=0$ in view of $|\lambda|<1$. Thus, solutions with constant $\lambda$ must necessarily involve a non-trivial $SU(1,1)$ monodromy on $\cA_\pm$. The existence of such solutions will be investigated in detail in a subsequent publication.

\sm

Allowing all three locally holomorphic functions $\cA_\pm$ and $\lambda$ to have non-trivial monodromy under $SU(1,1) \times \CC$, the mathematical problem becomes quite interesting, and quite involved. The pair  $(\cA_+,\cA_-)$ may then be viewed as a holomorphic section of a holomorphic bundle over $\Sigma$ with a structure group  which is a subgroup of $SU(1,1)\times \CC$. The fibers of this bundle are subject to the regularity conditions $\kappa ^2 >0$ and $\cG >0$.

\sm

The simplest situation is when $\Sigma$ is compact and without boundary, and  we shall attempt to give a plausible mathematical context for this case. The $SU(1,1) \times \CC$ invariance of $\cG$ then guarantees the continuity of $\cG$ as a function on $\Sigma$ and therefore its boundedness. Using the freedom to shift $\cB$ by an arbitrary constant, we may always adjust the shift to make $\cG >0$, and therefore to render this condition trivially satisfied. The only remaining condition  on the holomorphic sections $(\cA_+, \cA_-)$ of the bundle is then a condition on its associated one-forms,
\bea
|\p_w \cA_+|^2 - |\p_w \cA_-|^2 <0
\eea
The construction of this bundle should be expected to parallel the construction of vector bundles over $\Sigma$ with structure group contained in $SU(1,1)$, with the important difference that in our case the space of one-forms $\p_w \cA_\pm$ is subject to the above inequality which makes each fiber into a solid cone in $\CC^2$ rather than a vector space.

\sm

Holomorphic vector bundles over compact Riemann surfaces were considered early on  in \cite{Weil} and classified when their structure group is a subgroup of $SU(n)$ and the bundle satisfies certain stability conditions \cite{NS1,NS2}.\footnote{A useful introduction to this work, accessible to physicists, may be found in \cite{david}. We thank David Gieseker for pointing us to his paper.} The equivalence classes of rank $n$ holomorphic vector bundles are then in one-to-one correspondence with the irreducible representations of the fundamental homotopy group $\pi_1 (\Sigma)$ into $SU(n)$. The concrete realization of these representations on a compact surface of genus $g \geq 2$ is constructed as follows. We begin by introducing a basis of $A_i$ and $B_i$ cycles  for the first homology group $H_1 (\Sigma, \ZZ)$, with  canonical normalization of their intersection matrix $\#(A_i, A_j) = \# (B_i, B_j)=0$ and $\# (A_i, B_j)=\delta _{ij}$ for $i,j=1,\cdots, g$. The representation $\gamma$ of $\pi_1(\Sigma)$ into $SU(n)$ may then be concretely described by assigning elements $\gamma (A_i)$ and $\gamma (B_i)$ in $SU(n)$ to the homology cycles $A_i$ and $ B_i$ respectively, subject to the standard closure condition on the commutators,
\bea
\label{close}
\prod _{i=1}^g \gamma (A_i) \gamma (B_i) \gamma (A_i)^{-1} \gamma (B_i)^{-1} = I
\eea
The product in (\ref{close})  is ordered in the index $i$, and $I$ stands for the identity matrix in $SU(n)$. The discrete group  $\Gamma$ is then freely generated by the elements $\gamma (A_i)$ and $ \gamma (B_i)$ for $i=1,\cdots, g$, subject to the closure relation (\ref{close}).

\sm

Holomorphic vector bundles over $\Sigma$ whose structure group $\Gamma$ is a subgroup of $SL(2,\RR)=SU(1,1)$ instead should admit an analogous construction. We should assign to each homology generator $A_i, B_i$  a transformation $\gamma (A_i), \gamma (B_i)$ in $\Gamma$ subject to the closure relations  of (\ref{close}). The corresponding group $\Gamma$  is then freely generated by these elements, just as was the case for the unitary groups. In the simplest case where the structure group is $\ZZ_2$, the corresponding differential forms span the space of Prym differentials \cite{Fay} on the surface $\Sigma$, familiar from $\ZZ_2$ orbifold constructions in string theory. The case of a more general structure group $\Gamma$ may be viewed as a generalization of Prym differentials to the case of un-ramified covers of $\Sigma$ with higher structure group $\Gamma$. Finally, for our set-up the fibers are solid cones of one-forms  in $\CC^2$ rather than vector spaces, but the construction of such bundles is expected to follow in parallel.  A detailed investigation into these possibilities will be relegated to future work.

%%%%%%%%%%%%%%%%%%%%%%%%%%%%%%%%%%%%%%%%%%%%
%%%%%%%%%%%%%%%%%%%%%%%%%%%%%%%%%%%%%%%%%%%%
\section{Discussion}
%%%%%%%%%%%%%%%%%%%%%%%%%%%%%%%%%%%%%%%%%%%%
%%%%%%%%%%%%%%%%%%%%%%%%%%%%%%%%%%%%%%%%%%%%

Supersymmetric solutions to Type IIB supergravity with an $AdS_6$ factor and the corresponding isometries are of considerable interest to describe holographically the still mysterious 4+1 dimensional CFTs discovered in \cite{Seiberg:1996bd,Intriligator:1997pq} and realized in string theory via $(p,q)$ brane webs.
As quantum field theories these CFTs are defined only indirectly, as the non-trivial UV fixed points of certain 4+1 dimensional super-Yang-Mills theories. The description as Yang-Mills theories, however, is non-renormalizable and only captures the IR effective action of a relevant deformation of the CFT. Having supergravity duals at our disposal would allow for extensive quantitative studies of the actual UV CFTs, which from the field theory side are hard to do even in principle without a known Lagrangian description.

\sm

In this work we have constructed the general local solution to Type IIB supergravity with $SO(2,5) \times SO(3)$ isometry and 16 supersymmetries, which are precisely the symmetries expected for holographic duals to these 4+1 dimensional SCFTs.
The local solutions are constructed from $AdS_6$ and S$^2$ spaces, which realize the desired isometries and are fibered over a Riemann surface $\Sigma$.
With the supergravity fields spelled out in section~\ref{sec:2}, the local solutions are given in terms of two locally holomorphic functions $\cA_\pm$, as summarized in section~\ref{sec:5-1} and \ref{sec:5-2}.
The solutions transform properly under $SL(2,\RR)$, and this symmetry serves as an organizing guide throughout the derivations. Finally,  we recover the (singular) T-dual of the D4/D8 solution in Type IIA as a special case.

\sm

A crucial ingredient which has allowed us to go beyond earlier works, where the BPS equations were reduced to a set of coupled PDEs \cite{Apruzzi:2014qva,Kim:2015hya,Kim:2016rhs},
was keeping the freedom to choose convenient coordinates on the Riemann surface $\Sigma$.
Separating holomorphic and anti-holomorphic dependences in conformally flat coordinates on $\Sigma$ featured prominently in the derivations and allowed us to actually solve the reduced BPS equations.
The complete local solution provides the basic building blocks for constructing globally regular solutions with the aforementioned isometries, and is a significant step towards understanding and eventually classifying supergravity duals for the 4+1 dimensional CFTs realized by $(p,q)$ brane webs.

\sm

The natural next question is whether or not there are globally regular solutions with the symmetries discussed above.
While the existence of a large-$N$ limit on the CFT side suggests that there should be dual supergravity solutions,
the difficulties in finding such solutions so far call for a more careful perspective.
Indeed, there are known examples where a superconformal field theory with a large-$N$ limit does not admit a dual description which reduces to supergravity alone, even at large $N$ and strong coupling \cite{Gadde:2009dj}.
The existence of a supergravity dual is obstructed in that case by the absence of a gap in the spectrum to isolate a small number of states with low scaling dimension. As a result, the dual description always involves stringy states.
In the example of \cite{Gadde:2009dj}, however, realizing a superconformal theory at large $N$ needs a similarly large number of flavors, of the same order as $N$.
This Veneziano limit is different from the usual 't Hooft limit and from the case we are looking at here, where superconformal theories exist also with small numbers of flavors.
There is, quite on the contrary, an upper limit on the number of flavor multiplets in the CFTs we are attempting to find a dual description for.
The large-$N$ limit should therefore indeed be a 't~Hooft limit,
and these arguments suggest that finding a dual supergravity description should be possible.
We discussed regularity conditions in section~\ref{sec:5}, but constructing globally regular solutions is still a non-trivial task, and we plan to come back to it in the future.

\section*{Acknowledgements}
We are happy to acknowledge useful conversations with Oren Bergman and David Gieseker.
The work of EDH and MG is supported in part by the National Science Foundation under grant PHY-13-13986.
The work of MG is in part supported by the Simons Foundation through a Simons fellowship.
The work of AK and CFU was supported, in part, by the US Department of Energy under grant number DE-SC001163. MG is grateful for the Burke Center of Theoretical Physics, Caltech for hospitality during the course of this work.

\appendix

%%%%%%%%%%%%%%%%%%%%%%%%%%%%%%%%%%%%%%%%%%%
%%%%%%%%%%%%%%%%%%%%%%%%%%%%%%%%%%%%%%%%%%%
\section{Clifford algebra basis adapted to the Ansatz}
\setcounter{equation}{0}
\label{appA}
%%%%%%%%%%%%%%%%%%%%%%%%%%%%%%%%%%%%%%%%%%%
%%%%%%%%%%%%%%%%%%%%%%%%%%%%%%%%%%%%%%%%%%%

The signature of the space-time metric is chosen to be $(- + \cdots +)$. The Dirac-Clifford algebra is defined by $\{ \Gamma ^M, \Gamma ^N \} = 2 \eta ^{MN}  I_{32}$. We choose a basis for the Clifford algebra which is well-adapted to the $AdS_6 \times S^2  \times \Sigma$ Ansatz, with the frame labeled as in (\ref{frame1}),
\bea
\G^m & = & \g^m \otimes I_2 \otimes I_2  \hskip 1.08in m =0,1,2,3,4,5
\no \\
\G^{i} & = & \g_{(1)} \otimes \g^{i} \otimes I_2  \hskip 1.12in i=6,7
\no \\
\G^a \, & = & \g_{(1)}  \otimes \sigma ^3 \otimes  \gamma ^a \hskip 1.05in a=8,9
\eea
where  a convenient basis for the lower dimensional Dirac-Clifford algebras is as follows,
\bea
 \g^0 & =& -i \sigma ^2 \otimes I_2 \otimes I_2
\no \\
\g^1 &=& \sigma ^1 \otimes I_2 \otimes I_2
\no \\
\g^2 &=& \sigma ^3 \otimes \sigma ^2 \otimes I_2 
\hskip 1in \g^6 = \sigma^1
\no \\
\g^3 &=& \sigma ^3 \otimes \sigma ^1\otimes I_2
\hskip 1in \g^7 = \sigma^2
\no \\
\g^4 &=& \sigma ^3 \otimes \sigma ^3 \otimes \sigma ^1
\hskip 1in \g^8 = \sigma^1
\no \\
\g^5 &=& \sigma ^3 \otimes \sigma ^3\otimes  \sigma ^2
\hskip 1in \g^9 = \sigma^2
\eea
We will also need the chirality matrices on the various components of
$AdS_6 \times S^2  \times \Sigma$, and they are chosen as follows,
\bea
\gamma _{(1)} & = & \sigma ^3 \otimes \sigma ^3 \otimes \sigma ^3
\no \\
\gamma _{(2)} & = & \sigma ^3
\no \\
\gamma _{(3)} & = & \sigma ^3
\eea
so that,
\bea
\Gamma ^{012345} & = & - \gamma _{(1)} \otimes I_2 \otimes I_2
\no \\
\Gamma ^{67} & = & i \, I_6 \otimes \gamma _{(2)} \otimes I_2
\no \\
\Gamma ^{89} & = & i \, I_6 \otimes I_2 \otimes \gamma _{(3)}
\eea
The 10-dimensional chirality matrix in this basis is given by,
\bea
\G^{11} = \G^{0123456789}   = \g_{(1)} \otimes \g_{(2)} \otimes \g_{(3)}=\sigma ^3\otimes \sigma ^3\otimes \sigma ^3\otimes \sigma ^3\otimes \sigma ^3
\eea
The complex conjugation matrices in each component are defined by,
\bea
\label{bmatdef}
\left ( \g^m \right ) ^* = +B_{(1)} \g ^m B_{(1)} ^{-1}
& \hskip .5in &   (B_{(1)})^* B_{(1)} = - I_6 \hskip .6in B_{(1)} = -i \g^2 \gamma ^5=I_2 \otimes \sigma_1 \otimes \sigma_2
\no \\
\left ( \g^{i} \right ) ^* = -B_{(2)} \g ^{i} B_{(2)} ^{-1}
& &   (B_{(2)})^* B_{(2)} = -I_2 \hskip .6in B_{(2)} =  \g^7=\sigma^2
\no \\
\left ( \g^a \right ) ^* = - B_{(3)} \g ^a B_{(3)} ^{-1}
& &   (B_{(3)})^* B_{(3)} = - I_2 \hskip .6in B_{(3)} =  \g^9 =\sigma^2
\eea
where in the last column we have also listed the form of these matrices in our particular basis. The 10-dimensional complex conjugation matrix $\cB$ satisfies,
\bea
( \Gamma^M)^* = \cB \Gamma^M\cB^{-1}
\hskip 1in 
\cB ^* \cB=I_{32} 
\hskip 1in 
\ [ \cB, \Gamma^{11} \ ] = 0
\eea and in this basis is given by,
\bea
\cB &=&  -i B_{(1)}\otimes \left ( B_{(2)} \gamma _{(2)} \right ) \otimes B_{(3)}   \no\\
    & = &  \, I_2 \otimes \sigma ^1 \otimes \sigma ^2 \otimes \sigma ^1 \otimes \sigma ^2
\eea

%%%%%%%%%%%%%%%%%%%%%%%%%%%%%%%%%%%%%%%%%%%
%%%%%%%%%%%%%%%%%%%%%%%%%%%%%%%%%%%%%%%%%%%
\section{The geometry of Killing spinors}
\setcounter{equation}{0}
\label{appB}
%%%%%%%%%%%%%%%%%%%%%%%%%%%%%%%%%%%%%%%%%%%
%%%%%%%%%%%%%%%%%%%%%%%%%%%%%%%%%%%%%%%%%%%

In this Appendix, we review the relation between the Killing spinor
equation and the parallel transport equation in the presence of a flat connection
with torsion on $S^2$ and $AdS_4$.

\subsection{The sphere \texorpdfstring{$S^2$}{S2}}

On the sphere $S^2$ with unit radius, the Killing spinor equation is given by,
\bea
\label{KS1}
\left ( \nabla _i  +\eta  \half \sigma ^i \sigma ^3 \right )\ep & = & 0 \hskip 1in  i=1,2
\eea
where $\sigma ^1, \sigma ^2, \sigma ^3$ are the standard Pauli matrices, and $\nabla_i$ is the spin connection on the sphere $S^2$ with unit radius, and the derivative is expressed with respect to frame indices $i$. An equivalent equation is obtained by letting $\sigma ^i \sigma ^3 \to i \sigma ^i$ and $\ep ' = e^{- i \pi \sigma ^3/4} \ep$. Integrability of this system of equations requires $\eta = \pm 1$. The system may be solved by restricting the canonical flat connection on $SU(2)$, as was reviewed in an appendix of \cite{D'Hoker:2007xy}. Here, we will take a more direct approach to obtaining an explicit solution.

\sm

We parametrize the sphere $S^2$ with unit radius by angles $\theta _1, \theta _2$ and take the following conventions for the round metric,
\bea
\label{mets2}
ds^2=d\theta_2^2+\sin^2 \theta_2 d\theta_1^2
\eea
The frame and the spin connection are given by,
\bea
e^1= \sin\theta_2 \;d\theta_1 \qquad \qquad 
e^2= d\theta_2 \qquad \qquad
\omega^{12}=\cos\theta_2 \; d\theta_1
\eea
and satisfy the vanishing torsion condition,
\bea
de^a+ \omega^a{} _b \wedge e^b=0
\eea
The Killing spinor equation on the sphere $S^2$ of unit radius is of the form,
\bea
\Big(\partial_\mu+{1\over 4} \omega_\mu^{ab}\gamma_{ab}\Big)\ep = \eta {i\over 2} e^a_\mu \gamma_a \ep
\eea
with $\eta=\pm 1$ required  by integrability. With the Dirac matrices defined in appendix \ref{appA}  we have $\gamma^1=\sigma_1$ and $\gamma^2=\sigma_2$, and  this equation becomes,
\bea
\partial_{\theta_1} \ep +{i\over 2} \cos \theta_2 \sigma_3 \, \ep = 
{i\over 2 }  \eta \, \sin \theta_2 \, \sigma_1 \, \ep 
\hskip 1in 
\partial_{\theta_2}  \ep = {i\over 2} \eta \, \sigma_2 \, \ep 
\label{killing2}
\eea
and is solved by,
\bea
\ep (\eta)= \exp \left ( {i\over2 } \eta \theta_2 \sigma_2 \right ) 
\exp \left ( -{i\over 2} \theta_1 \sigma_3 \right ) \, \ep_0
\eea
Here, $\ep_0$ is an arbitrary constant spinor. Hence the space of Killing spinors is two-dimensional. We can verify the following statements for a Killing spinor $\ep(\eta)$ which satisfies the Killing equation with a given $\eta=\pm1$,
\begin{enumerate}
\itemsep=0in
\item $\gamma_{(2)} \ep(\eta) =\sigma_3 \ep(\eta)$ satisfies the Killing equation with $\eta\to -\eta$;
\item  $\sigma_1 \ep(\eta)^*$ satisfies the Killing equation with $\eta\to -\eta$;
\item $\sigma_2 \ep(\eta)^*$ satisfies the Killing equation with $\eta\to +\eta$.
\end{enumerate}

\subsection{Minkowski \texorpdfstring{$AdS_6$}{AdS6}}

The construction based on canonical connections given in \cite{D'Hoker:2007xy} for $S^2$ may be generalized to all spheres and their hyperbolic $AdS$ counterparts. Here, we present the case of Minkowski signature $AdS_6 = SO(2,5)/SO(1,5)$. The Dirac-Clifford algebra of $SO(2,5)$ is built from the Clifford generators $\gamma ^\mu$,  of the Lorentz group $SO(1,5)$,
\bea
\{ \gamma ^\mu , \gamma ^\nu\} = 2 \eta ^{\mu \nu }
\hskip 1in
\eta = {\rm diag} [-+++++]
\eea
for $\mu, \nu =0,1,2,3$, supplemented with the chirality matrix,
$\gamma ^{\sharp } = i \gamma _{(1)} $ (the factor of $i$ being required since the $\natural$
dimension has negative entry in the metric),
\bea
\{ \gamma ^ {\bar \mu} , \gamma ^{\bar \nu} \} = 2 \bar \eta ^{\bar \mu \bar \nu}
 \hskip 1in
\bar \eta = {\rm diag} [--+++++]
\eea
for $\bar \mu, \bar \nu = \sharp , 0,1,2,3,4,5$.
The corresponding Maurer-Cartan form, expressed in the spinor representation, on $SO(2,5)$ is given by,
\bea
\omega ^{(t)} = V ^{-1} dV = {1 \over 4} \omega ^{(t)} _{\bar \mu \bar \nu}
\gamma ^{\bar \mu \bar \nu} \hskip 1in V \in  Spin(2,5)
\eea
It obviously satisfies the Maurer-Cartan equations, $d \omega ^{(t)}
+ \omega ^{(t)} \wedge \omega ^{(t)}=0$.
We decompose $\omega ^{(t)}$ onto the $SO(1,5)$ and $AdS_6$
directions of cotangent space,
\bea
\omega ^{(t)} = { 1 \over 4} \omega _{\mu \nu} \gamma ^{\mu \nu}
+ {i \over 2} e_\mu \gamma ^\mu \gamma _{(1)}
\hskip 1in
\left\{
\begin{array}{cc}
\omega _{\mu \nu}   \equiv \omega ^{(t)} _{\mu \nu } & \mu,\nu =0,1,2,3,4,5 \\
e_\mu \equiv \omega ^{(t)} _{\mu \sharp} & \mu=0,1,2,3,4,5
\end{array} \right .
\eea
The Maurer-Cartan equations $ d \omega ^{(t)} + \omega ^{(t)} \wedge \omega ^{(t)}=0$
for $\omega ^{(t)}$ imply the absence of torsion and the constancy of curvature.
The Killing spinor equation coincides with the equation for parallel transport,
\bea
\left ( d + V^{-1} dV \right ) \ep =
\left ( d + { 1 \over 4} \omega _{\mu \nu} \gamma ^{\mu \nu}
+ {i \over 2} \eta e_\mu \gamma ^\mu \gamma _{(1)}  \right ) \ep =0
\eea
For $\eta=+1$, the general solution is given by $\ep _+ = V^{-1} \ep _0$ and $\ep_0$ is constant, while for $\eta =-1$, the solution is $\ep_- = \gamma _{(1)} \ep _+$. An equivalent form of the Killing spinor equation,  in a  basis in which the chirality matrix $\gamma _{(1)}$ is eliminated,  is achieved by making the transformation,
\bea
\ep  = \exp \left ( { i \pi \over 4} \gamma _{(1)} \right ) \ep '
\eea
In terms of $\ep'$, we recover the first equation in (\ref{3a1}).

\subsubsection{Explicit form of the Killing spinors}

We parametrize $AdS_6$ with unit radius by a radial coordinate $r$ and $x^i \in \RR$ for $i=1,2,3,4$, and use the $SO(2,5)$-invariant metric,
\bea
ds^2 = dr^2 + e^{2r} \left (-dt^2+\sum_{i=1}^4 d x_i^2 \right )
\eea
The Killing spinor equation is given by,
\bea
D_\mu \ep = {\eta \over 2} \gamma_\mu \ep
\eea
with $\eta=\pm1$. The frame and spin connection are given by,
\bea
 e^{ r} = dr \hskip 1in   e^{ i} = e^r dx^i  \hskip 1in \omega^{i r  }= e^{r} dx^i
\eea
Hence the Killing spinor equation becomes,
\bea
\label{killing6}
\partial_r \ep =  {\eta \over 2} \gamma_{ r}\ep
\hskip 1in 
\partial_{i} \ep+{1\over 2} e^r \gamma_{ i r} \ep = {\eta\over 2} e^r \gamma_{ i} \ep
\eea
The general solution for the Killing spinor equation is,
\bea
\ep = e^{{\eta\over 2} r \gamma_{ r} } \Big( 1+ {1\over 2} x^i \gamma_{i}(\eta - \gamma_{r}) \Big) \ep_0
\eea
Where $\ep_0$ is a constant spinor, hence the space of Killing spinors of $AdS_6$ is eight dimensional.  We can label the basis vectors by $\eta=\pm 1$ which each have a four dimensional degeneracy which will play no role in the following other than leading to the correct number of preserved supersymmetries in the end.
We can verify the following statements for an $AdS_6$ Killing spinor $\epsilon(\eta)$ which satisfies the Killing equation with a given $\eta=\pm1$,
\begin{enumerate}
\itemsep=0in
\item $\gamma_{(1)} \epsilon(\eta) =\sigma_3\otimes \sigma_3\otimes \sigma_3 \; \epsilon(\eta)$ satisfies the Killing equation with $\eta\to -\eta$;
\item  $B_{(1)} \epsilon(\eta)^* = 1_2\otimes \sigma_2\otimes\sigma_1 \;  \epsilon(\eta)^*$ satisfies the Killing equation with $\eta\to \eta$;
\item $\tilde B_{(1)} \epsilon(\eta)^* = \sigma_3 \otimes \sigma_1\otimes \sigma_2 \;\epsilon(\eta)^*$ satisfies the Killing equation with $\eta\to -\eta$.
\end{enumerate}
Note that we choose the complex conjugation matrix to be $B_{(1)}$ in  appendix \ref{appA}

%%%%%%%%%%%%%%%%%%%%%%%%%%%%%%%%%%%%%%%%%%%
%%%%%%%%%%%%%%%%%%%%%%%%%%%%%%%%%%%%%%%%%%%
\section{Derivation of the BPS equations}
\setcounter{equation}{0}
\label{appC}
%%%%%%%%%%%%%%%%%%%%%%%%%%%%%%%%%%%%%%%%%%%
%%%%%%%%%%%%%%%%%%%%%%%%%%%%%%%%%%%%%%%%%%%

We begin by collecting some identities that will be useful during the  reduction of the BPS equations. We will also need the following decompositions of $\ep$ and $\cB ^{-1} \ep ^*$,
\bea
\ep = \sum_{\eta_1, \eta_2, \eta_3} \chi^{\eta_1, \eta_2} \otimes
\zeta_{\eta_1, \eta_2}
\hskip 1in 
\cB^{-1} \varepsilon^* = \sum_{\eta_1, \eta_2} \chi^{\eta_1, \eta_2} \otimes
\star \zeta _{\eta _1, \eta _2}
\eea
where we use the abbreviations,
\bea
\label{star}
\star \zeta _{\eta _1, \eta _2}
=  - i \sigma ^2  \eta _2  \zeta _{\eta _1,-\eta _2}^*
\hskip 1in 
\star \zeta =  \tau ^{02} \otimes \sigma ^2 \zeta ^*
\eea
in $\tau$-matrix notation. In addition, we have the chirality relations,
\bea
\sigma ^3 \zeta _{\eta _1, \eta _2} = - \zeta _{-\eta_1, - \eta _2}
\hskip 1in
\tau ^{11} \otimes \sigma ^3 \zeta = - \zeta
\eea

%%%%%%%%%%%%%%%%%%%%%%%%%%%%%%%%%%%%%%%%%%%
\subsection{The dilatino equation}
%%%%%%%%%%%%%%%%%%%%%%%%%%%%%%%%%%%%%%%%%%%

The dilatino equation is,
\bea
0=  i P_A  \G^A \cB^{-1} \ep ^* -{i\over 24} \G \cdot  G \ep
\eea
Reduced to the Ansatz of subsection \ref{sec:2-3}, we have the following simplifications,
\bea
P_A \G^A &=&p_a  \G^a
\no\\
\G \cdot  G &=& 3! \, g_a \G^{67a} = 6  g_a  \Gamma ^a \, i \,  I_8 \otimes \gamma _{(2)}  \otimes I_2
\eea
The dilatino equation now becomes,
\bea
0 = i p_a \G^a \sum_{\eta_1 ,\eta_2} \chi^{\eta_1 ,\eta_2} \otimes
\star \zeta_{\eta_1, \eta_2}
+ {1 \over 4} g_a \Gamma ^a  \sum_{\eta_1, \eta_2}
 \chi^{\eta_1, - \eta_2} \otimes \zeta_{\eta_1, \eta_2}
\eea
from which we extract the equation satisfied by the $\zeta$-spinors,
\bea
0 = i p_a \gamma^a
\star \zeta_{\eta_1, \eta_2}
+ {1 \over 4} g_a \gamma ^a   \zeta_{\eta_1, - \eta_2}
\eea
Using the explicit expression for $\star \zeta$ and  reversing the sign of $\eta_2$, we find,
\bea
0 = - p_a \gamma^a
\eta _2 \sigma ^2  \zeta_{\eta_1, \eta_2}^*
+ {1 \over 4} g_a \gamma ^a   \zeta_{\eta_1,  \eta_2}
\eea
Recasting this equation in terms of the $\tau$-notations, we have,
\bea
0 = - 4 p_a \tau^{(03)} \gamma ^a \sigma ^2 \zeta ^* +  g_a \gamma ^a \zeta
\eea

%%%%%%%%%%%%%%%%%%%%%%%%%%%%%%%%%%%%%%%%%%%
\subsection{The gravitino equation}
%%%%%%%%%%%%%%%%%%%%%%%%%%%%%%%%%%%%%%%%%%%

The gravitino equation is,
\bea
0 & = & d \ep + \omega \ep - {i \over 2} Q  \ep + \mg \,  \cB^{-1} \ep^*
\no \\
\omega &=& {1 \over 4} \omega_{AB} \Gamma^{AB}
\no\\
\mg &=& - {1 \over 96} e_A \bigg( \G^A (\G \cdot G) + 2 (\G \cdot G) \G^A \bigg)
\eea
where $A,B$ are the 10-dimensional frame indices.

\subsubsection{The calculation of \texorpdfstring{$(d + \omega) \ep$}{the derivative}}

The spin connection components are $\omega ^a {}_b$, whose explicit form we will not need and,
\bea
\label{spincon}
\omega^m {}_n = \hat \omega^m {}_n
&\qquad&
\omega^m {}_a = e^m \, {\p_a f_6 \over f_6}
\no\\
\omega^i {}_j = \hat \omega^i {}_j ~
&\qquad&
\omega^i {}_a = e^i \, {\p_a f_2 \over f_2}
\eea
The hats refer to the canonical connections on $AdS_6$ and $S^2$ respectively.
Projecting the spin-connection along the various directions we have,
\bea
(m) &\qquad& \nabla_m^\prime \ep + {D_a f_6 \over 2 f_6} \, \G_{m} \, \G^{a} \, \ep
\no\\
(i) &\qquad& \nabla_i ^\prime \, \ep +  {D_a f_2 \over 2 f_2} \, \G_i \, \G^{a} \, \ep
\no\\
(a) &\qquad& \nabla_a \ep
\eea
where the prime on the covariant derivative indicates that only the connection  along $AdS_6$ and  $S^2$  respectively is included.   Using the Killing spinor equations (\ref{3a1}) we can eliminate the primed covariant derivatives,  which yields,
\bea
(m) &\qquad& {1 \over 2 f_6} \G_{m}
\sum_{\eta_1 ,\eta_2 } \eta_1 \chi^{\eta_1, \eta_2} \otimes
\zeta_{\eta_1, \eta_2} + {D_a f_6 \over 2 f_6} \, \G_{m} \G^{a} \, \ep
\no\\
(i) &\qquad& {i \over 2f_2} \G_i
\sum_{\eta_1, \eta_2} \eta_2 \chi^{- \eta_1, \eta_2} \otimes
\zeta_{\eta_1, \eta_2} + {D_a f_2 \over 2 f_2} \G_i \G^{a} \ep
\eea
Using the equation $\G^a = \gamma_{(1)} \otimes \gamma_{(2)} \otimes \gamma_{(3)} \otimes \gamma^a$, we have,
\bea
(m) &\qquad& \G_m \sum_{\eta_1, \eta_2} \chi^{\eta_1, \eta_2} \otimes
\bigg( {1 \over 2 f_6} \eta_1 \zeta_{\eta_1, \eta_2}
+ {D_a f_6 \over 2 f_6} \, \gamma^{a} \zeta_{- \eta_1, - \eta_2} \bigg)
\no\\
(i) &\qquad& \G_i \sum_{\eta_1 ,\eta_2} \chi^{\eta_1, \eta_2} \otimes
\bigg( {i \over 2 f_2} \eta_2 \zeta_{- \eta_1, \eta_2}
+ {D_a f_2 \over 2 f_2} \gamma^{a} \zeta_{-\eta_1, -\eta_2} \bigg)
\eea
where we have pulled a factor of $\G_M$ out front.  It will turn out that  all terms in the gravitino equation contain $\G_M \chi^{\eta_1, \eta_2}$,  and we will require the coefficients to vanish independently, just as we did for  the dilatino equation.  The coefficient of  $\G_M \chi^{\eta_1 ,\eta_2}$ can be expressed in the $\tau$-matrix  notation as,
\bea
\label{redomega}
(m) &\qquad& {1 \over 2 f_6} \tau^{(30)}  \zeta
+ {D_a f_6 \over 2 f_6} \, \tau^{(11)}  \gamma^{a} \zeta
\no\\
(i) &\qquad& {i \over 2 f_2} \tau^{(13)}  \zeta
+ {D_a f_2 \over 2 f_2} \tau^{(11)}  \gamma^{a} \zeta
\eea

\subsubsection{The calculation of \texorpdfstring{$\mg \,  \cB^{-1} \ep^*$}{charge conjugate term}}

The relevant expression is as follows,
\bea
\mg \,  \cB^{-1} \ep^* = - {3! \over 96} e_B g_a \left ( \G^B \G^{67a} + 2 \G^{67a} \G^B \right )
 \cB^{-1} \ep^*
\eea
A few useful equations are as follows,
\bea
\G^m \G^{67b} + 2 \G^{67b} \G^m &=& - \G^m \G^{67b}
\no \\
\G^i \, \G^{67b} + 2 \G^{67b} \, \G^i &=& 3 \G^i \, \G^{67b}
\no \\
\G^a \G^{67b} + 2 \G^{67b} \G^a &=& \G^{67} (3 \delta^{ab} - \G^{ab})
= i (I_8 \otimes \gamma_{(2)} \otimes I_2) (3 \delta^{ab}I_2 -  \gamma^{ab})
\eea
where $\gamma ^{ab} = \ep^{ab} \sigma ^3$ and $\ep ^{89}=1$.
Projecting along the various directions we obtain,
\bea
(m) &\quad& \G_m \sum_{\eta_1, \eta_2} \chi^{\eta_1 ,\eta_2}
\otimes \left (  {i \over 16} \, g_a \gamma^{a} \star  \zeta_{- \eta_1, \eta_2}  \right )
\no \\
(i) &\quad&   \G_i \,  \sum_{\eta_1, \eta_2} \chi^{\eta_1, \eta_2}
\otimes \left (  - {3 i \over 16} \, g_a \gamma^{a} \star \zeta_{- \eta_1, \eta_2} \right )
 \no\\
(a) &\quad& \sum_{\eta_1, \eta_2} \chi^{\eta_1, \eta_2}
\otimes \bigg( - {3 i \over 16}   g_a \star \zeta_{\eta_1, - \eta_2}
+ { i \over 16}  g_b \gamma^{ab} \star \zeta_{\eta_1, - \eta_2}  \bigg)
\eea
Using the $\tau$-matrix notation, we can write the coefficient of
$\G_M \chi^{\eta_1, \eta_2}$ in the form,
\bea
\label{redg}
(m) &&
+ {i \over 16} \, g_a \tau^{(10)}  \gamma^{a} \star \zeta
\no \\
(i) &&
- {3i \over 16} \, g_a \tau^{(10)} \gamma^{a} \star \zeta
\no\\
(a) &&
- {3i \over 16}  g_a \tau^{(01)}  \star \zeta
+ {i \over 16} g_b \tau^{(01)}  \gamma^{ab} \star \zeta
\eea

\subsubsection{Assembling the complete gravitino BPS equation}

Now we combine the reduced gravitino equations.   We again argue that the $\Gamma_M \chi^{\eta_1 \eta_2}$  are linearly independent which leads to the equations,
\bea
(m) &\qquad&
0 = {1 \over 2 f_6} \tau^{(30)}  \zeta
+ {D_a f_6 \over 2 f_6} \tau^{(11)}  \gamma^{a} \zeta
+ {i \over 16}  g_a \tau^{(10)}  \gamma^{a} \star \zeta
\\
(i) &\qquad&
0 = {i \over 2 f_2} \tau^{(13)}  \zeta
+ {D_a f_2 \over 2 f_2} \tau^{(11)}  \gamma^{a} \zeta
- {3 i \over 16}  g_a \tau^{(10)} \gamma^{a} \star \zeta
\no\\
(a) &\qquad&
0 = \bigg( D_a  + {i \over 2} \hat \omega _a   \sigma^{3} - { i \over 2} q_a \bigg) \zeta
- {3 i \over 16} g_a \tau^{(01)} \star \zeta + {i \over 16}  g_b \tau^{(01)}  \gamma^{ab} \star \zeta
\no
\eea
where $\hat \omega _a = (\hat \omega _{89})_a$ is the spin connection along $\Sigma$. Eliminating the star using the definition $(\ref{star})$,  namely $\star \zeta =  \tau^{(02)} \otimes \sigma^2 \zeta^*$.  The system of gravitino BPS  equations is then,
\bea
(m) &\qquad&
0 = {1 \over 2 f_6} \tau^{(30)}  \zeta
+ {D_a f_6 \over 2 f_6} \tau^{(11)}  \gamma^{a} \zeta
+ {i \over 16}  g_a \tau^{(12)}  \gamma^{a}  \sigma^2 \zeta^*
\\
(i) &\qquad&
0 = {i \over 2 f_2} \tau^{(13)}  \zeta
+ {D_a f_2 \over 2 f_2} \tau^{(11)}  \gamma^{a} \zeta
- {3 i \over 16}  g_a \tau^{(12)} \gamma^{a}  \sigma^2 \zeta^*
\no\\
(a) &\qquad&
0 = \bigg( D_a  + {i \over 2} \hat \omega _a   \sigma^{3} - { i \over 2} q_a \bigg) \zeta
+ {3  \over 16} g_a \tau^{(03)}   \sigma^2 \zeta^*
- {1 \over 16}  g_b \tau^{(03)}  \gamma^{ab}  \sigma^2 \zeta^*
\no
\eea
Upon multiplying the $(m)$ and $(i)$ equations by $ \tau ^{(11)}$, we find,
\bea
(m) &\qquad&
0 = - {i \over  2f_6} \tau^{(21)}  \zeta
+ {D_a f_6 \over  2f_6}   \gamma^{a} \zeta
- {1 \over 16}  g_a \tau^{(03)}  \gamma^{a}  \sigma^2 \zeta^*
\\
(i) &\qquad&
0 = {1 \over  2 f_2} \tau^{(02)}  \zeta
+ {D_a f_2 \over  2f_2}   \gamma^{a} \zeta
+ {3  \over 16}  g_a \tau^{(03)} \gamma^{a}  \sigma^2 \zeta^*
\no\\
(a) &\qquad&
0 = \bigg( D_a  + {i \over 2} \hat \omega _a   \sigma^{3} - { i \over 2} q_a \bigg) \zeta
+ {3  \over 16} g_a \tau^{(03)}   \sigma^2 \zeta^*
- {1 \over 16}  g_b \tau^{(03)}  \gamma^{ab}  \sigma^2 \zeta^*
\no
\eea

%%%%%%%%%%%%%%%%%%%%%%%%%%%%%%%%%%%%%%%%%%%%%%%%%%%%%%
%%%%%%%%%%%%%%%%%%%%%%%%%%%%%%%%%%%%%%%%%%%%%%%%%%%%%%
\section{Verifying the Bianchi identities}
\setcounter{equation}{0}
\label{app:Bianchi-G}
%%%%%%%%%%%%%%%%%%%%%%%%%%%%%%%%%%%%%%%%%%%%%%%%%%%%%%
%%%%%%%%%%%%%%%%%%%%%%%%%%%%%%%%%%%%%%%%%%%%%%%%%%%%%%

In this appendix we show that the reduced BPS equations summarized in section~\ref{sec:4-2} imply that the Bianchi identities (\ref{bianchi1}) are satisfied. In the complex coordinates introduced in (\ref{eq:w-coords}), the Bianchi identities for $P$ and $Q$ take the following form,
\bea
\label{6a4}
\p_w P_{\bar w} - \pbw P_w - 2 i Q_w P_{\bar w} + 2 i Q_{\bar w} P_w & = & 0
\no \\
\p_w Q_{\bar w} - \pbw Q_w +  i P_w \bar P_{\bar w} -  i P_{\bar w} \bar P_w & = & 0
\eea
They are satisfied automatically if $P$ and $Q$ are expressed in terms of $B$ as in (\ref{sugra1}).
The Bianchi identity that needs to be checked is the one for $G$.
We will find that (\ref{5c5}) alone implies that it is satisfied.
In terms of the lower-case expansion coefficients $p$, $q$, $g$ defined in (\ref{PQdef}),
the Bianchi identity (\ref{bianchi1}) becomes,
\begin{equation}
 \partial_w\big(\rho f_2^2g_{\bar z}\big)
 -\partial_{\overline{w}}\big(\rho f_2^2g_z\big)
 +\rho^2f_2^2\big(-iq_z g_{\bar z}+iq_{\bar z}g_z+p_z(g_z)^\star-p_{\bar z}(g_{\bar z})^\star\big)
\end{equation}
Using (\ref{5a1}) and (\ref{sugra1}), this can be rewritten in terms of $B$, $\rho$ and the $\kappa_\pm$ alone
as,
\begin{align}
\label{eqn:reduced-Bianchi-G}
 \Big(\frac{a^2}{b^2}+e^{-i\vartheta}\Big)\left[\bigtriangleup B+\frac{5}{2}f^2\bar{B} (\partial_w B) (\partial_{\overline{w}}B)\right]
 +\frac{a^2}{b^2}(\partial_w B)\partial_{\overline{w}}\log\Big[\frac{a}{b}\frac{(a\bar a-b\bar b)^2}{\rho^2}\Big]\hspace*{15mm}
 \no\\
 +e^{-i\vartheta}(\partial_{\overline{w}}B)\partial_w \log\Big[\frac{\bar b}{\bar a}\frac{(a\bar a-b\bar b)^2}{\rho^2}\Big]
 +\Big(\frac{3}{2}B+e^{i\vartheta}\Big)f^2\left(\frac{a^2}{b^2}|\partial_w B|^2+e^{-i\vartheta}|\partial_{\overline{w}}B|^2\right)=0
\end{align}
where $\bigtriangleup=\partial_w\partial_{\overline{w}}$ and,
\bea
\label{eqn:ab}
\bar a^2 & = &  \rho \bar\alpha^2/f=\kappa_++B\kappa_-
\no \\
\bar b^2 & = & \rho\bar \beta^2/f=\bar{B}\kappa_++\kappa_-
\eea
are used as shorthands for the corresponding combinations of  $\kappa_\pm$. The only place where $\rho^2$ shows up is in the log derivatives,  where it can be eliminated using (\ref{5c5}), leaving an expression solely in terms of $B$ and $\bar B$.

\sm

To show that (\ref{5c5}) implies that the Bianchi identity is satisfied, we first eliminate the two-derivative term.
To this end, we act on (\ref{eqn:Beq}) with $\partial_w$
and on the complex conjugate of (\ref{eqn:Beq}) with $\partial_{\overline{w}}$.
The result are two equations linear in $\bigtriangleup B$ and $\bigtriangleup \bar B$, namely,
\begin{eqnarray}\no
 \bar X\bigtriangleup B -\bar Y\bigtriangleup \bar B+ (\partial_w B)\partial_{\overline{w}} \bar X
 -(\partial_w \bar B)\partial_{\overline{w}}\bar Y
 &=&
 (\partial_{\overline{w}}f^{-4})\kappa_+^2\partial_w\frac{\kappa_-}{\kappa_+}
 \\
 X\bigtriangleup \bar B-Y\bigtriangleup B+(\partial_{\overline{w}} \bar B)\partial_w X-(\partial_{\overline{w}} B)\partial_w Y
 &=&
 (\partial_w f^{-4})\bar\kappa_+^2\partial_{\overline{w}}\frac{\bar\kappa_-}{\bar\kappa_+}
\end{eqnarray}
where $X=b^4+2a^2b^2e^{i\theta}$ and $Y=a^4+2a^2b^2e^{-i\theta}$.
Eliminating $\bigtriangleup \bar B$ yields an equation for $\bigtriangleup B$ in terms of first derivatives, namely,
\begin{eqnarray}
\no
 \bigtriangleup B \big(|X|^2-|Y|^2\big)&=&
 X(\partial_w \bar B)\partial_{\overline{w}}\bar Y
 +\bar Y(\partial_{\overline{w}}B)\partial_w Y
 -X(\partial_w B)\partial_{\overline{w}}\bar X
 -\bar Y(\partial_{\overline{w}}\bar B)\partial_w X
 \\
 &&
 +X(\partial_{\overline{w}}f^{-4})\kappa_+^2\partial_w\frac{\kappa_-}{\kappa_+}
 +\bar{Y}(\partial_w f^{-4})\bar\kappa_+^2\partial_{\overline{w}}\frac{\bar\kappa_-}{\bar\kappa_+}
\end{eqnarray}
Substituting this expression in (\ref{eqn:reduced-Bianchi-G}) leaves only first derivatives.
We can now use (\ref{eqn:Beq}) and its complex conjugate to eliminate the derivatives of $\bar B$.
After this step the left hand side of (\ref{eqn:reduced-Bianchi-G}) collapses to zero, showing that the Bianchi identity is satisfied.

%%%%%%%%%%%%%%%%%%%%%%%%%%%%%%%%%%%%%%%%%%%
%%%%%%%%%%%%%%%%%%%%%%%%%%%%%%%%%%%%%%%%%%%
\section{Supergravity fields in terms of holomorphic data}
\setcounter{equation}{0}
\label{sec:10}
%%%%%%%%%%%%%%%%%%%%%%%%%%%%%%%%%%%%%%%%%%%
%%%%%%%%%%%%%%%%%%%%%%%%%%%%%%%%%%%%%%%%%%%

In this appendix we derive the expression for the supergravity fields from the local solution to the BPS equations derived in section~\ref{sec:4}. The results are summarized in section~\ref{sec:5}. As a prerequisite, we note that (\ref{eq:5-2a}) arises from (\ref{X0}) with,
\begin{align}
 \kappa_-\bar\cL &=-\partial_w\cG
 &
\kappa^2&=-\partial_w\partial_{\bar{w}}\cG
\end{align}
We then start with the metric functions and the axion-dilaton combination, before turning to the flux field.

\subsection{Metric functions}

We start with the expressions for the metric functions $f_2$, $f_6$ in terms of $\alpha$, $\beta$ given in (\ref{f2f6-2}).
Using the expressions for $\alpha$, $\beta$ in terms of $\kappa_\pm$ and $f$ given in (\ref{5b4}) yields
\begin{equation}
 \Big(f_6+\frac{3}{\nu} f_2\Big)^2=
 \frac{4c_6^2 f^2}{\rho^2}|\kappa_-|^2|B\bar\lambda+1|^2
 \qquad\quad
 \Big(f_6-\frac{3}{\nu} f_2\Big)^2=
 \frac{4c_6^2 f^2}{\rho^2}|\kappa_-|^2 |\lambda+B|^2
\end{equation}
Translating to $Z$ and using (\ref{8a3}) for $f^2$ yields
\begin{equation}
f_6+\frac{3}{\nu} f_2=
\frac{2c_6}{\rho}\sqrt{\frac{\kappa^2}{1-|Z|^4}}
\qquad\quad
f_6-\frac{3}{\nu} f_2=\frac{2c_6}{\rho}|Z|^2\sqrt{\frac{\kappa^2}{1-|Z|^4}}
\end{equation}
Solving for $f_2$ and $f_6$ yields
\begin{equation}
\label{f2f6-Z}
 f_2=\frac{\nu c_6}{3 \rho}\sqrt{\kappa^2\frac{1-|Z|^2}{1+|Z|^2}}
 \qquad\qquad
 f_6=\frac{c_6}{\rho}\sqrt{\kappa^2\frac{1+|Z|^2}{1-|Z|^2}}
\end{equation}
With $|Z|^2=R$ and $\nu^2=1$, this immediately leads to the expressions in (\ref{eq:5-2b}).
Next comes $\rho^2$, the metric on $\Sigma$. From (\ref{8d1}), (\ref{8e4}) and (\ref{calA}),
\begin{equation}
 \hat\rho^2=\frac{\rho^2}{c_6|\kappa_-|^2}\frac{(1-|Z|^2)^{3/2}}{|Z|\sqrt{1-|\lambda|^2}\sqrt{1+|Z|^2}}
 \qquad\quad
 \hat\rho^4=\frac{1}{\xi\bar\xi}=\frac{(1-\lambda\bar\lambda)^2}{\cL \bar\cL}
\end{equation}
From those one finds
\begin{equation}\label{rho2-Z}
 \rho^2=
 \frac{c_6\kappa^2}{|\kappa_-\bar \cL|}\, \frac{|Z|}{1-|Z|^2}
 \sqrt{\kappa^2\frac{1+|Z|^2}{1-|Z|^2}}
\end{equation}
With $|Z|^2=R$ and $\kappa_-\bar\cL=-\partial_w\cG$, this directly leads to the expression quoted in (\ref{eq:5-2b}).

\sm

To get to the axion and dilaton we work out the expression for $B$.
Using the definition in (\ref{8a1}) and (\ref{8d1})
\begin{equation}
\label{9B1}
 B=\frac{R e^{i\psi}-\lambda}{1-\bar\lambda R e^{i\psi}}
 =\frac{R\bar\cL -\lambda \cL }{\cL -\bar\lambda R\bar\cL }
\end{equation}
The second equality uses $e^{i\psi}=\bar\xi/\xi$ and (\ref{calA}).
Multiplying numerator and denominator by $|\kappa_-|^2$ yields
\begin{equation}
\label{9B2}
 B=\frac{\bar\kappa_-(R\kappa_-\bar\cL )-\kappa_+(\bar\kappa_-\cL)}
 {\kappa_-(\bar\kappa_-\cL)-\bar\kappa_+(R\kappa_-\bar\cL)}~.
\end{equation}
The explicit expression for $f^2$ can be obtained from (\ref{8a3}) as,
\begin{equation}\label{9f}
 f^2=
 \frac{|\kappa_-|^2}{\kappa^2|\cL|^2}\frac{|\bar\cL -\lambda R\cL|^2}{1-R^2}~.
\end{equation}
For $\tau$ we find
\begin{equation}
 i\tau=\frac{\lambda_+-\bar\lambda_+Z^2}{\lambda_-+\bar\lambda_-Z^2}
 =\frac{\lambda_+\cL -\bar\lambda_+\bar\cL R}{\lambda_-\cL +\bar\lambda_-\bar\cL R}
 \qquad\quad
 \lambda_\pm=\lambda\pm 1
\end{equation}
To get to the axion and dilaton we separate the real and imaginary parts
and use $\tau=\chi+ie^{-2\phi}$, which yields
\begin{equation}
\label{9phi}
 e^{-2\phi}=\frac{\kappa^2|\cL |^2}{D|\kappa_-|^2}\frac{1-R^2}{R}
 \qquad \qquad
 D=|\lambda_-\cL |^2 W+\lambda_-^2\cL^2+\bar\lambda_-^2\bar\cL^2
\end{equation}
The axion becomes
\begin{equation}\label{9chi}
 \chi=i\frac{(\lambda-\bar\lambda)|\cL|^2 W-\lambda_+\lambda_-\cL^2 +\bar\lambda_+\bar\lambda_-\bar\cL^2}{D}
\end{equation}

\subsection{Flux field}

The complex  3-from field $G_{(3)}$ is better expressed in terms of the complex 3-form field $F_{(3)}$ by the first relation of (\ref{GF5}). Since $F_{(3)}$ is a closed 3-form it may locally  be written in terms of a complex flux potential 2-form field $\CB$ by $F_{(3)} = d \CB$. Inverting this relation, we have,
\bea
d\CB = f (G_{(3)} + B \bar G_{(3)})
\eea
The symmetries of the problem force $\CB$ and  $G_{(3)}$ to be of the following form,
\bea
\CB & = & \cC \, \hat e^{67}
\no \\
G_{(3)} & = & g_a e^a \wedge e^{67} = f_2^2 \, g_a e^a\wedge \hat e^{67}
\eea
so that,
\bea
\p_w \cC & = & \rho f_2^2 f ( g_z + B \bar g_z)
\no \\
\pbw \cC & = & \rho f_2^2 f ( g_{\bar z} + B \bar g _{\bar z} )
\eea
Using the conversion of $G$ into $P$ and then into derivatives of $B$ using (\ref{sugra1}) and (\ref{5a1}),
\bea
\rho \, g_z = 4i {\alpha \over \beta } f^2 \p_w B & \hskip 1in &
\rho \, g_{\bar z} = - 4i { \bar \beta \over \bar \alpha} f^2 \pbw B
\no \\
\rho \, \bar g_z = 4i {\beta \over \alpha } f^2 \p_w \bar B & \hskip 1in &
\rho \, \bar g_{\bar z} = - 4i { \bar \alpha \over \bar \beta } f^2 \pbw \bar B
\eea
we obtain the following expressions,
\bea
\p_w \cC & = &  4 i f_2^2 f^3 \left ( { \alpha \over \beta} \p_w B + B { \beta \over \alpha} \p_w \bar B \right )
\no \\
\p_w \bar \cC & = &  4 i f_2^2 f^3 \left ( {  \beta  \over  \alpha } \p_w \bar B + \bar B {  \alpha  \over  \beta } \p_w B \right )
\eea
We will now work towards expressing the right side in terms of the solutions to the BPS equations,
and then integrating the equations to obtain $\cC$ and its complex conjugate.

\subsubsection{Expressing variables in terms of holomorphic functions}

Recall that we have,
\bea
{ \alpha \over \beta} =
\left ( { \bar \lambda + \bar B \over \bar \lambda B +1} \right )^\half
=
\bar Z \left ( { 1 - \bar \lambda Z^2 \over 1 - \lambda \bar Z^2} \right )^\half
\hskip 1in
\left | { \alpha \over \beta} \right | = |Z|
\eea
This allows us to compute the combinations,
\bea
{ \alpha ^2 \over \beta ^2 } \p_w B + B \pbw \bar B
& = &
{ \bar Z^2 (1 - |\lambda|^2) \over |1-\lambda \bar Z^2|^2} \p_w Z^2
+ { (Z^2-\lambda )(1-|\lambda|^2) \over (1-\lambda \bar Z^2) |1-\lambda \bar Z^2|^2} \p_w \bar Z^2
\no \\ &&
- { \bar Z^2 (1 - |\lambda|^2) ( 1 - |Z|^4)   \over (1-\lambda \bar Z^2) |1-\lambda \bar Z^2|^2} \,
\p_w \lambda
\no \\
{\beta ^2 \over \alpha ^2} \p_w \bar B + \bar B \p_w B
& = & { 1 - |\lambda|^2 \over |1-\lambda \bar Z^2|^2} \left ( { \p_w \bar Z^2 \over \bar Z^2} +{ \bar Z^2 - \bar \lambda \over 1 - \bar \lambda Z^2} \p_w Z^2 \right )
\eea
In the second equation, the terms proportional to $\p_w \lambda$ cancel. Putting all together, we have,
\bea
\p_w \cC  & =  &  { 4i c_6 \over 9 \hat \rho^2}
{ (1 -|Z|^2) \over   (1 +|Z|^2)^3} \, { 1 \over \bar Z |Z|}
\Big (
 \p_w |Z|^4 - \lambda (  \p_w \bar Z^2 + \bar Z^4 \p_w Z^2 ) - \bar Z^2 (1-|Z|^4) \p_w \lambda  \Big )
 \no \\
 \p_w \bar \cC  & =  &  { 4i c_6 \over 9 \hat \rho^2}
{ (1 -|Z|^2) \over   (1 +|Z|^2)^3} \, { 1 \over \bar Z |Z|}
\Big ( - \bar \lambda  \p_w |Z|^4  +  \p_w \bar Z^2 +  \bar Z^4 \p_w Z^2     \Big )
\eea
Next, we eliminate $\hat \rho^2$ in favor of $\xi$, by using the following rearrangement formula,
\bea
{ 1 \over \hat \rho^2 \, \bar Z |Z|} = { \bar \xi \over |Z|^2}
\eea
so that we find,
\bea
\p_w \cC  & =  & { 4i c_6 \over 9 }
{ (1 -|Z|^2) \over   (1 +|Z|^2)^3} \, { \bar \xi \over  |Z|^2}
\Big (
 \p_w |Z|^4 - \lambda (  \p_w \bar Z^2 + \bar Z^4 \p_w Z^2 ) - \bar Z^2 (1-|Z|^4) \p_w \lambda  \Big )
 \no \\
 \p_w \bar \cC  & =  &  { 4i c_6 \over 9 }
{ (1 -|Z|^2) \over   (1 +|Z|^2)^3} \, { \bar \xi \over  |Z|^2}
\Big ( - \bar \lambda  \p_w |Z|^4  +  \p_w \bar Z^2 +  \bar Z^4 \p_w Z^2     \Big )
\eea
Instead of working with $\cC$ and $\bar \cC$, we form the  combination $\bar \cC + \bar \lambda \cC$,
\bea
\p_w (\bar \cC + \bar \lambda \cC) &=&  { 4 i c_6 \over 9} \, \cP
\eea
where $\cP$ is given by,
\bea
\cP  =  \xi \, { 1-  |Z|^2  \over (1+|Z|^2)^3} \Big ( (1-|\lambda|^2 ) (\p_w \ln \bar Z^2 + \bar Z^2 \p_w Z^2 )
+ (1-|Z|^4) \p_w ( 1- |\lambda|^2) \Big )
\eea
Having $\bar \cC + \bar \lambda \cC$ is essentially as good as having $\cC$ itself, since from $\bar \cC + \bar \lambda \cC$ and its complex conjugate we may obtain $\cC$ and $\bar \cC$.
Next, we use the formula $\xi (1-|\lambda|^2) = \cL$ to eliminate $\xi$ in favor of $\lambda$ and $\cL$,
\bea
\cP  = \cL { 1-  |Z|^2  \over (1+|Z|^2)^3} \Big ( \p_w \ln \bar Z^2 + \bar Z^2 \p_w Z^2
+ (1-|Z|^4) \p_w \ln ( 1- |\lambda|^2) \Big )
\eea
Changing variables from $Z$ to $Z^2 = R e^{i \psi}$, we find,
\bea
\cP  = \cL   {  (1-R)(R^2+1) \over (1+R)^3} {\p_w R \over R}  
- \cL {(R-1)^2 \over (R+1)^2}  \Big ( i \p_w \psi -
\p_w \ln ( 1- |\lambda|^2) \Big )
\eea
Expressing this combination in terms of $W=R+R^{-1}$, we find,
\bea
\cP  = - \cL   {W \over (W+2)^2} {\p_w W}  - \cL {W-2 \over W+2}  \Big ( i \p_w \psi -
\p_w \ln ( 1- |\lambda|^2) \Big )
\eea
Using equation (\ref{Xeq}), we eliminate  $\hat \rho$ in favor of $\xi$, and eliminate $\xi$ 
in favor of $\cL$ and $\lambda$,
\bea
\label{pW}
\p_w W =  - (W+1) \p_w \ln  \cL \bar \cL  +(W-2) \p_w \ln (1- |\lambda|^2) + 3 i \p_w \psi
\eea
Putting all together, we have,
\bea
\cP & = & { \cL \over (W+2)^2} \Big (  W(W+1) \p_w \ln \cL \bar \cL + 2(W-2) \p_w \ln ( 1 - |\lambda|^2)
\no \\ && \hskip 1in
 - (W^2+3W-4) \p_w \ln { \bar \cL \over \cL} \Big )
\eea
where we have used the relation $e^{ i \psi} = \bar \cL / \cL$ to express the derivative $i \p_w \psi$ in terms of $\cL$. Finally, $W$ is given in terms of holomorphic data by,
\bea
W = 2 + 6 \, { 1 - |\lambda|^2 \over \cL \bar \cL} \, \Big ( \cA_+ \bar \cA_+ - \cA_- \bar \cA_- + \cB + \bar \cB \Big )
\eea
Separating the dependence of $\cL$ and $\bar \cL$, we find,
\bea
\label{LW}
\cP  = { 2 \cL \over (W+2)^2} \Big (  (W^2+2W-2)  \p_w \ln \cL   + (W-2) \left \{ \p_w \ln ( 1 - |\lambda|^2) - \p_w \ln \bar \cL \right \}  \Big )
\eea
Next, we return to the differential equation satisfied by $W$ in (\ref{pW}) and notice that the precise same
combination $(W-2) \p_w \ln (1-|\lambda|^2)$ occurs there. Expressing $\psi$ in terms of $\cL$ in this equation, it takes the form,
\bea
(W-2) \left \{ \p_w \ln ( 1 - |\lambda|^2) - \p_w \ln \bar \cL \right \}
= \p_w W +(W+4) \p_w \ln \cL
\eea
Thus, we will proceed to eliminate this term between (\ref{pW}) and (\ref{LW}), and we find,
\bea
\cP = { 2 (W+1)  \over W+2} \p_w \cL  + { 2  \cL  \over (W+2)^2} \, \p_w W
\eea
It is straightforward to integrate this equation, and we have,
\bea
\cP = \p_w \left ( { 2 \cL \, (W+1) \over W+2} \right )
\eea
Therefore, we have,
\bea
\bar \cC + \bar \lambda \cC = \bar \cK_1 + { 8 i c_6 \over 9} \,  {  \cL \, (W+1) \over W+2}
\eea
for an as yet undetermined holomorphic function $\cK_1$.

\sm

Proceeding analogously for $\bar \cC$, and using the equations for $\p_w \xi$ and $\p_w \bar \xi$, we find,
\bea
\p_ w \bar \cC = { 4 i c_6 \over 9} \p_w \left ( - 2 { \xi + \bar \lambda \bar \xi \over W+2} 
+ { \cL \over 1 - |\lambda|^2} + \cA_- \right )
\eea
so that
\bea
\bar \cC = { 4 i c_6 \over 9} \left ( - 2 { \xi + \bar \lambda \bar \xi \over W+2} 
+ { \cL \over 1 - |\lambda|^2} + \cA_- + \bar \cK_2 \right )
\eea
for some holomorphic function $\cK_2$. Equating now the two different expressions for $\bar \cC + \bar \lambda \cC$, we get after some simplifications which eliminate the dependence on $W$ completely,
\bea
\bar \cK_1 - \bar \cA_+ - \bar \cK_2 + \bar \lambda ( 3 \bar \cA_- - 2 \cA_+ + \cK_2)=0
\eea
Separating holomorphic and anti-holomorphic dependences, we find, 
\bea
\cK_1 = 3 \cA_+ - 3 \lambda \cA_- + \cK_0 + \bar \cK_0 \lambda
\eea
where $\cK_0$ is an arbitrary complex constant.  Thus, we have,
\bea
\label{eq:C}
\cC = { 4 i c_6 \over 9} \left ( + 2 \, { \bar \xi +  \lambda  \xi \over W+2} - { \bar \cL \over 1 - |\lambda|^2} - \bar  \cA_- - 2 \cA_+ - \cK_0 \right )
\eea
Evaluating that further gives
\bea
\cC = { 4 i c_6 \over 9} \left (  {  2\lambda  \cL - W \bar \cL \over (W+2)(1-|\lambda|^2) }  - \bar  \cA_- - 2 \cA_+ - \cK_0 \right )
\eea

\newpage

\bibliographystyle{JHEP}
\bibliography{ads6}
\end{document}